\documentclass[journal]{IEEEtran}
\usepackage{amsmath,amsfonts}
\usepackage{algorithmic}
\usepackage[linesnumbered,ruled]{algorithm2e}
\usepackage{array}
\usepackage{textcomp}
\usepackage{stfloats}
\usepackage{url}
\usepackage{verbatim}
\usepackage{graphicx}
\usepackage{subfigure}
\usepackage{cite}
\usepackage{multirow}
\usepackage{xcolor}
\usepackage{placeins}
\begin{document}

\title{Digital Twin Based User-Centric Resource Management for Multicast Short Video Streaming}

\author{\IEEEauthorblockN{Xinyu Huang, \textit{Student Member, IEEE},  Wen Wu, \textit{Senior Member, IEEE}, Shisheng Hu, \textit{Student Member, IEEE}, Mushu Li, \textit{Member, IEEE}, Conghao Zhou, \textit{Member, IEEE}, and Xuemin (Sherman) Shen, \textit{Fellow, IEEE}
		}

\thanks{Xinyu Huang, Shisheng Hu, Conghao Zhou, and Xuemin (Sherman) Shen are with
	the Department of Electrical and Computer Engineering, University of
	Waterloo, Waterloo, ON N2L 3G1, Canada (e-mail:
	\{357huan, s97hu, c89zhou, sshen\}@uwaterloo.ca);\\ \indent
	Wen Wu is with the Frontier Research Center, Peng Cheng Laboratory, Shenzhen 518055, China (e-mail: wuw02@pcl.ac.cn); \\ \indent
Mushu Li is with the Department of Electrical, Computer and Biomedical Engineering, Toronto Metropolitan University, Toronto, ON M5B 2K3,
Canada (e-mail: mushu.li@ieee.org).}}



\maketitle
\nocite{prework}
\begin{abstract}

Multicast short video streaming (MSVS) can effectively reduce network traffic load by delivering identical video sequences to multiple users simultaneously. The existing MSVS schemes mainly rely on the aggregated video requests to reserve bandwidth and computing resources, which cannot satisfy users’ diverse and dynamic service requirements, particularly when users’ swipe behaviors exhibit spatiotemporal fluctuation. In this paper, we propose a user-centric resource management scheme based on the digital twin (DT) technique, which aims to enhance user satisfaction as well as reduce resource consumption. Firstly, we design a user DT (UDT)-assisted resource reservation framework. Specifically, UDTs are constructed for individual users, which store users' historical data for updating multicast groups and abstracting useful information. The swipe probability distributions and recommended video lists are abstracted from UDTs to predict bandwidth and computing resource demands. Parameterized sigmoid functions are leveraged to characterize multicast groups’ user satisfaction. Secondly, we formulate a joint non-convex bandwidth and computing resource reservation problem which is transformed into a convex piecewise problem by utilizing a tangent function to approximately substitute the concave part. A low-complexity scheduling algorithm is then developed to find the optimal resource reservation decisions. Simulation results based on the real-world dataset demonstrate that the proposed scheme outperforms benchmark schemes in terms of user satisfaction and resource consumption.
\end{abstract}

\begin{IEEEkeywords}
Digital twin, resource management, multicast transmission, user-centric.
\end{IEEEkeywords}

\section{Introduction}
The proliferation of short video platforms, such as TikTok, Instagram Reels, and YouTube Shorts, enabled by the ubiquity of smartphones and high-speed wireless networks, has ushered in a new era of digital entertainment \cite{yuan2019spatial, wang2022task}. According to a recent report from TikTok, the number of active global users has risen from 902 million to 1.47 billion in 2022 and will continue to maintain significant growth \cite{report}. However, this short video streaming surge puts a significant burden on the existing wireless network infrastructures. Particularly, the ultra-definition and panoramic short videos poses new requirements on the transmission and computing capabilities of wireless networks, especially on higher bandwidth (300 Mbps) and transcoding speed (30.2~FPS), for providing immersive user experience \cite{panor, li2023utility}. Multicast transmission, as an important technology in wireless networks, allows a single data stream to be disseminated to numerous users in a group simultaneously. By applying multicast transmission to short video streaming, the bandwidth utilization and network throughput can be effectively enhanced\cite{Multicast, 9676649}. 

To support multicast short video streaming (MSVS), network resources, including bandwidth and computing resources, are required. Specifically, spectrum bandwidths are required for video delivery to each multicast group \cite{yans}, and computing resources are required for video transcoding for each multicast group \cite{xinyu_transcoding}. For consistent quality of service (QoS) provisioning, such network resources need to be reserved in advance \cite{reservation},\cite{hebo}. Nevertheless, existing resource reservation schemes are mainly based on the aggregated video requests while neglecting the unique users' behaviors in watching short videos, i.e., \textit{swipe behaviors}. Without considering the swiping behaviors that can make videos not completely played by users, the resource demands can be overestimated and the reserved resources can be underutilized. As the swipe behaviors are user-specific \cite{zhu2022swipe}, resource reservation can incorporate the user-specific swipe behaviors, i.e., \textit{user-centric resource reservation}, is desired for efficient resource utilization in the MSVS.

Digital twin (DT) is a potential technology to realize the user-centric resource reservation. DT is defined as a full digital representation of a physical object, and real-time synchronization between the physical object and its corresponding digital replica \cite{digital}. To update current end users' statuses, such as network conditions, data traffic, and mobility trajectories, \textit{user DTs} (UDTs) are constructed for individual users. The data stored in UDTs can be leveraged to analyze users’ transmission rates and behavior patterns for resource reservation. In the MSVS scenario, UDTs can store users' historical networking and personal information, and analyze their video traffic patterns and watching behaviors. Based on the distilled coarse- and fine-grained user information from UDTs, users’ bandwidth and computing resource demands can be accurately predicted to facilitate the effective user-centric resource reservation.

Designing an effective UDT-based resource reservation scheme needs to address the following challenges: 1) incorporating swipe behaviors in resource reservation decision-making; 2) establishing the mathematical model between multicast groups’ user satisfaction and reserved resources. Specifically, users’ swipe behaviors are stochastic and spatiotemporally varied, which are difficult to be predicted in real time. Therefore, how to conduct effective data abstraction to obtain the distilled swipe feature and utilize it to predict bandwidth and computing resource demands is challenging. Furthermore, due to the dynamics of users’ personalized preferences and sensitivities to service quality, the same amount of reserved resources to one multicast group can lead to different user satisfaction at different time. Therefore, how to establish and update each multicast group's user satisfaction model in each resource reservation window is challenging.

In this paper, we \textit{firstly} propose a novel UDT-assisted resource reservation framework to incorporate the impact of swipe behaviors. Specifically, we construct UDTs to store users’ historical data, including channel conditions, locations, swipe timestamps, and preferences. A novel user clustering algorithm is proposed to analyze UDTs’ data for multicast group construction. The proposed user clustering algorithm consists of three parts, i.e., autoencoders, a double deep Q-network (DDQN), and the K-means++ method, which are responsible for UDTs’ data compression, clustering number determination, and fast user clustering, respectively. Then, the group-level information, i.e., the swipe probability distribution and recommended video list, for each multicast group is abstracted. Based on that, multicast groups’ average engagement time, video traffic, and computing consumption can be analyzed to predict bandwidth and computing resource demands for resource reservation. \textit{Secondly}, we establish a user satisfaction model to quantify the impact of bandwidth and computing resource reservation on user satisfaction. Specifically, we adopt a real-world user satisfaction dataset to explore the relationship between user satisfaction and reserved resources. Based on data fitting, we build two kinds of parameterized sigmoid functions to characterize the exponential relationship. \textit{Thirdly}, our objective is to maximize the system utility consisting of user satisfaction and resource consumption in each resource reservation window. Since the formulated problem is non-convex and difficult to solve, a low-complexity resource scheduling algorithm based on the linear approximation method to make the optimal resource reservation decisions is proposed. The extensive simulation results on real-world short video streaming datasets show that the proposed scheme can effectively improve system utility as compared with the state-of-the-art resource reservation schemes. 

The main contributions are summarized as follows:

\begin{itemize}
	
	\item[$\bullet$] We propose a novel UDT-assisted resource reservation framework for resource demand prediction;
	
	\item[$\bullet$] We establish a user satisfaction model to quantify the impact of reserved resources on user satisfaction;
	
	\item[$\bullet$] We propose a low-complexity scheduling algorithm to  determine bandwidth and computing resource reservation.
\end{itemize}

The remainder of this paper is organized as follows. Related works are presented in Section II. The considered scenario and the UDT-assisted resource reservation scheme are presented in Sections III and IV, respectively. The user satisfaction, the problem formulation, and the proposed scheduling algorithm are presented in Sections V, VI and VII, respectively. Simulation results are provided in Section VIII. Finally, Section IX concludes this paper.

\section{Related Work}
MSVS can distribute video sequences from a single base station (BS) to multiple users concurrently over the same wireless channels. This approach exhibits two primary advantages, i.e., efficient bandwidth utilization and scalable user scale \cite{xu2019optimal},\cite{li2018performance}. To facilitate the MSVS within 5G networks, extensive works are devoted to optimizing multicast transmission performance from different directions, such as the novel architecture design by integrating network function virtualization and mobile edge computing technique \cite{zahoor2020service}, transmission orchestration by leveraging collected global network information \cite{Zhang-mult, 10077734}, and signal multiplexing with non-orthogonal multiple access to transmit different segment layers \cite{NOMA-group}. These studies demonstrate innovative approaches toward enhancing the performance of MSVS.

To ensure the service quality for the MSVS, resource reservation for multicast groups is essential. Historical video contents and bitrates are usually selected to forecast users’ future peak video traffic, which can provide a primary basis for bandwidth resource reservation \cite{soni2017nfv}. Additionally, bitrate fluctuation and video quality are usually leveraged to predict transcoding consumption, which can serve as a fundamental foundation for VM instance reservation \cite{qiao2018improving}. To flexibly adjust resource reservation in the high-mobility and ultra-density network scenario, existing hierarchical resource reservation schemes usually predict video traffic in grid-partitioned regions \cite{araniti2018hybrid}, estimate basic resources per region based on QoS requirements \cite{ye2018network}, and employ machine learning methods to fine-tune resource reservation \cite{xu2021tripres}. However, these schemes do not incorporate the dynamics of user behavior, such as swipe probability, on-off patterns, etc., potentially leading to inaccurate user clustering. Furthermore, solely relying on aggregated video requests to reflect regional traffic patterns is not enough to support accurate resource demand prediction. In this case, reserved resources can occur underutilization.

DT as an important virtualization technology was first introduced to monitor and mitigate anomalous events for flying vehicles \cite{glaessgen2012digital}. By introducing DTs into the next-generation wireless network, the existing network architecture can realize holistic network virtualization for flexible and accurate resource management. We refer readers to recent comprehensive surveys and tutorials on DTs to be familiar with this topic \cite{holi, 10183802, 9939166}. There also exist some technical papers aiming at utilizing DTs to improve network performance. Sun \textit{et al}. and Huynh \textit{et al}. utilized DTs to estimate edge servers’ states and the entire MEC system, which can provide seamless and accurate training data for offloading decisions \cite{sun2020reducing}\cite{van2022urllc}. To coordinate the computing resource management of edge servers in real time, DTs were used to timely monitor the centralized training process for resource scheduling in the aerial edge networks \cite{liu2023energy}. Furthermore, DTs were used to capture the time-varying demand of computing resources of IoT devices to assist the computation offloading decisions in \cite{peng2022distributed}. A learning-based prediction model residing in the DT was developed to predict the waiting time of relays and transmit the predicted results in  \cite{zheng2023data}, which can efficiently synchronize real-time data. Lu \textit{et al}. formulated an edge association problem for adaptively placing and migrating DTs based on the dynamics of network states and end users, which can efficiently reduce the service latency \cite{lu2021adaptive}. These pioneering works can effectively improve resource management performance in terms of resource allocation, edge association, and task offloading.

Recently, a few early research works focused on utilizing DTs to enhance network performance in terms of resource reservation. Zhou \textit{et al}. constructed UDTs for individual users to analyze their mobility patterns and proposed an improved particle swarm optimization method to enable customized resource reservation \cite{zhou2022}. To alleviate data sparsity when performing the DT-assisted traffic prediction, a short-term traffic flow and velocity prediction method based on data similarity was proposed in \cite{hu-traffic}. Furthermore, since it is difficult to operate DTs on different granular levels in a large and heterogeneous user scale, Vaezi \textit{et al}. proposed a five-stage implementation framework to progressively abstract component-level, subsystem-level, and global-level information \cite{vaezi2022digital}. As an extension of our previous work \cite{prework}, we further propose a user satisfaction model to quantify the impact of reserved resources, taking the dynamics of users’ personalized preferences and sensitivities to service quality. A fast resource reservation algorithm is also proposed to determine bandwidth and VM instance reservation.

\section{Considered Scenario}
\begin{figure}[t]
	\centering
	\includegraphics[width=6.6cm]{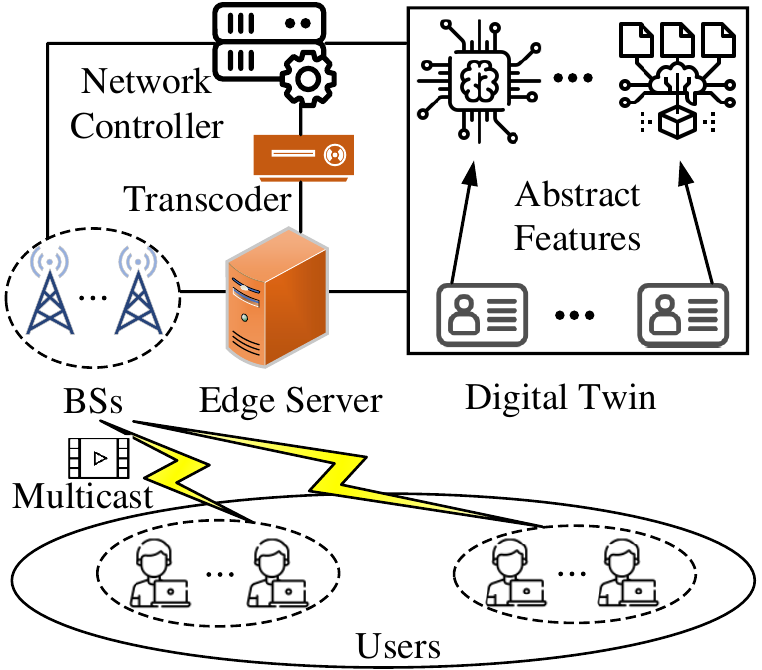}
	\caption{UDT-assisted resource reservation framework for MSVS.}
	\label{fig:framework}
\end{figure}

As shown in Fig.~\ref{fig:framework}, we consider a UDT-assisted MSVS scenario, which consists of multiple BSs, an edge server (ES), users, and UDTs.
\begin{itemize}
	
	\item[$\bullet$] BSs: BSs utilize multicast technology to transmit short videos to different multicast groups. The set of multicast groups is denoted by $\mathcal{G}=\left\{ 1,\cdots ,G \right\}$. The set of bandwidths for all BSs is denoted by $\mathcal{M}=\left\{ 1,\cdots ,M \right\}$.
	
	\item[$\bullet$] ES: The ES connects to all BSs and stores popular short videos with the highest version (bitrate) to avoid frequent video retrievals from content providers. The stored short videos can be transcoded to a lower version in the transcoder to adapt to users’ dynamic channel conditions and swipe behaviors. The set of VM instances for the transcoders is denoted by $\mathcal{N}=\left\{ 1,\cdots ,N \right\}$, and the computing capability of each VM instance is denoted by $\omega$. 
	
	\item[$\bullet$] Users: Users with similar statuses, such as preferences, swipe behaviors, locations, and channel conditions, are clustered into one multicast group, and the corresponding user set is denoted by ${{\mathcal{K}}_{g}}\text{=}\left\{ 1,\cdots ,{{K}_{g}} \right\}$. Users in the same multicast group are recommended the same short video list and receive them by the multicast transmission.
	
	\item[$\bullet$] UDTs: There are multiple UDTs deployed at the ES. Each UDT corresponds to a user and stores users’ statuses. BSs are responsible for collecting real-time users’ statuses to update UDTs. UDTs’ data are utilized to abstract some useful information, such as data similarity and swipe probability distribution, which can help the network controller realize an accurate multicast group construction and resource demand prediction.
\end{itemize}

The UDT-assisted resource reservation framework for MSVS operates as follows. In each resource reservation window, user statuses are uploaded to the ES to update corresponding UDTs and adjust multicast groups. In each multicast group, UDTs’ swipe timestamps and preferences are used to abstract the swipe probability distribution and recommended video list for accurate bandwidth and computing resource demand prediction. Based on the predicted information, the network controller reserves appropriate bandwidths and VM instances for each multicast group.

\section{UDT-assisted Resource Reservation Scheme}

\subsection{UDT Construction}
\begin{figure}[t]
	\centering
	\includegraphics[width=8cm]{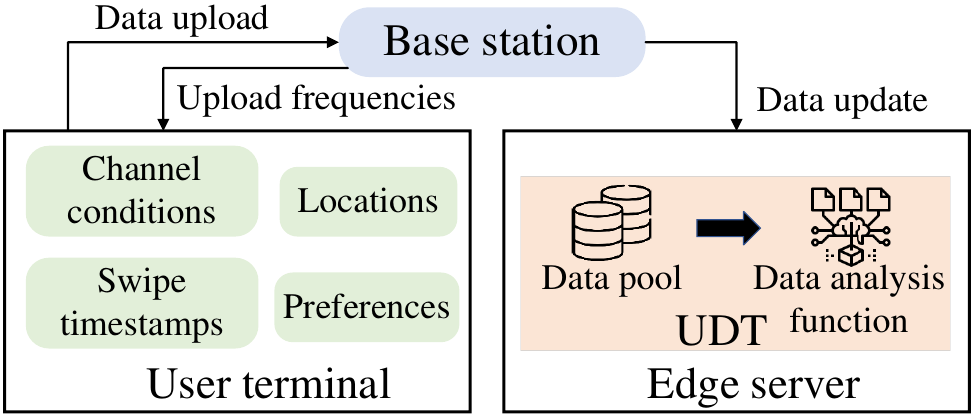}
	\caption{UDT data collection procedure.}
	\label{fig:UDT}
\end{figure}
UDT consists of a finite data pool and a data analysis function as shown in Fig.~\ref{fig:UDT}. The data pool records a user’s recent status through diversified data collection frequencies. The data analysis function investigates a user’s swipe timestamps to obtain a swipe probability distribution for each video type. The construction of a UDT can be summarized into two phases.
\subsubsection{Data Collection}
As shown in Fig.~\ref{fig:UDT}, BS collects a user’s data from two aspects, i.e., networking-related data and behavior-related data. The networking-related data include a user’s real-time channel conditions and locations, which are utilized to estimate the transmission rate. The behavior-related data consist of a user’s swipe timestamps and preferences. The swipe timestamps reflect a user's swipe behavior \cite{li2023dashlet}, which are used to abstract swipe probability distribution for engagement time prediction. The preferences are leveraged for the recommended video list update. By integrating these two kinds of data, UDTs can well reflect users' real-time statuses and predict bandwidth and computing resource demands.

Data collection number can be different for various UDT data attributes in each resource reservation window. Given that a low-speed mobility scenario where users’ channel conditions usually vary on a small timescale \cite{G_Huang}, the data collection number of a user’s channel condition and location in one resource reservation window, $T$, is denoted by $F_1$. Since the swipe probability needs to be calculated over several short videos, the corresponding data collection number of a user’s swipe timestamps in one resource reservation window is denoted by $F_2$. A user’s preference is an essential metric to determine which videos should be recommended. Since the user’s preference is relatively dynamic data that is updated based on the user’s like, share, and swipe frequency \cite{Gong}, the data collection number keeps consistent with that of the swipe timestamps to ensure the data freshness.

\subsubsection{Data Analysis}
A UDT’s functionality consists of two parts, i.e., a user’s swipe behavior analysis module and a data interaction module. The former analyzes the collected swipe timestamps to abstract swipe probability distributions. The latter provides stored data and the abstracted swipe probability distributions to the network controller to update multicast groups and predict resource demands. In resource reservation window $t$, user $k$’s swipe probability distribution, $p_{k,c}^{t}\left( e \right)$, for video type $c$ is updated by
\begin{equation}
p_{k,c}^{t}\left( e \right)=\lambda p_{k,c}^{t-1}\left( e \right)+\left( 1-\lambda  \right)\frac{A_{k,c}^{t-1}\left( e \right)}{\widehat{A}_{k,c}^{t-1}},	
\end{equation}
where $e$ is the video timestamp ranging from 1 to 15. Parameter $\lambda$ is a weighting factor ranging from 0 to 1. The higher $\lambda$ means a smaller impact of previous swipe behaviors on user $k$’s current swipe probability distribution. Here, $A_{k,c}^{t-1}\left( e \right)$ represents user $k$’s swipe numbers for video type $c$ in previous window $t-1$, which is counted by swipe timestamps. In addition, $\widehat{A}_{k,c}^{t-1}$ indicates the number of transmitted videos of type $c$ to user $k$ in previous window $t-1$. 

\subsection{UDT-Assisted Multicast Group}
Based on stored data in UDTs, we can analyze users' similarities to divide users with similar statuses into the same multicast group. A three-step method is proposed to realize fast and accurate multicast group construction. Specifically, autoencoders are first employed to compress time-series UDTs' data for dimension reduction. Then, a deep reinforcement learning (DRL) network is adopted to determine the clustering number by mining users’ intrinsic correlation. Finally, the K-means++ algorithm \cite{kmeans} is utilized to realize a fast multicast group construction based on the determined clustering number \cite{wenwen}. After multicast groups are constructed, we can abstract the swipe probability distribution and recommended video list for each multicast group.

\subsubsection{Clustering Number Determination} 
\begin{figure}[t]
	\centering
	\includegraphics[width=8.8cm]{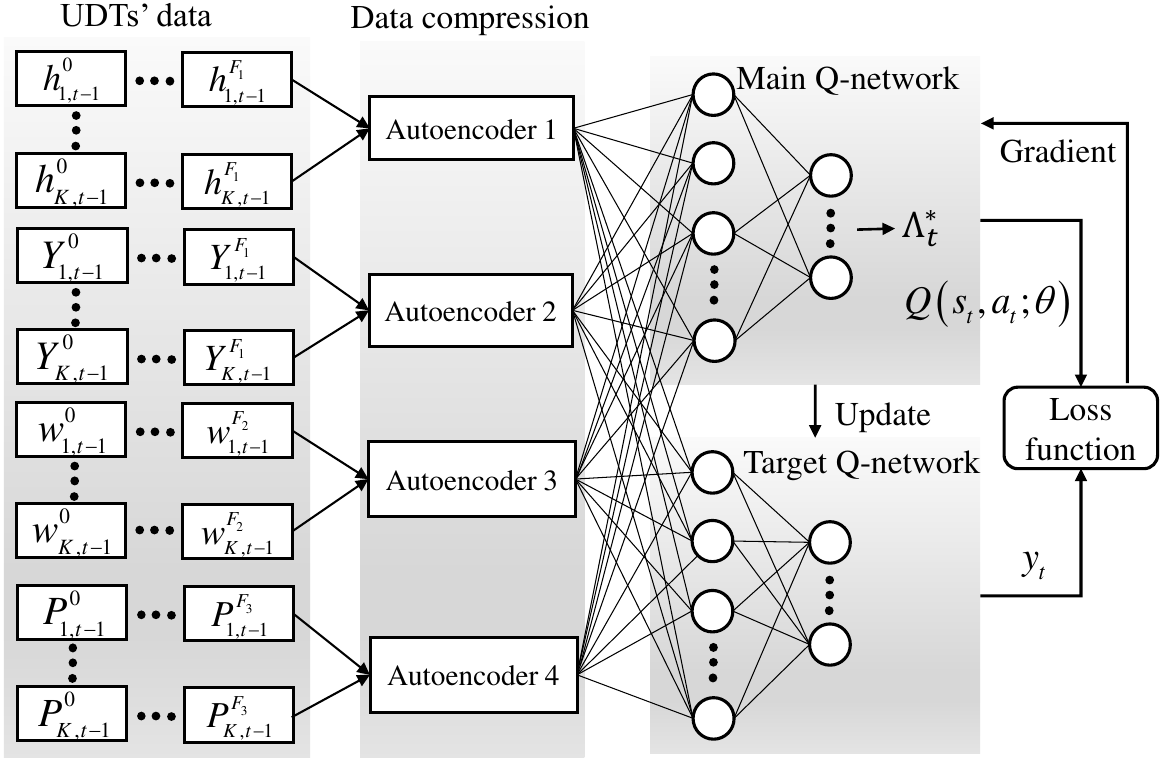}
	\caption{The improved DDQN architecture for clustering number determination.}
	\label{fig:reinforce}
\end{figure}

UDTs’ data consist of four dimensions and each dimension corresponds to time-series data, directly using a DRL method to analyze UDTs’ data may suffer from the curse of input dimension. Therefore, we add four Autoencoders to the existing DDQN \cite{DDQN} to compress UDTs' data, which can effectively reduce the input dimension \cite{Autoencoder}. The compressed data are further input to the actor-critic network for determining an appropriate clustering number, as shown in Fig.~\ref{fig:reinforce}. 

As shown in Fig.~\ref{fig:reinforce}, the processed UDTs’ data consist of users’ historical channel conditions, ${{\left\{ h_{k,t-1}^{f} \right\}}_{k\in \mathcal{K},f\in {{\mathcal{F}}_{1}}}}$, locations, ${{\left\{ Y_{k,t-1}^{f} \right\}}_{k\in \mathcal{K},f\in {{\mathcal{F}}_{1}}}}$, swipe timestamps, ${{\left\{ w_{k,t-1}^{f} \right\}}_{k\in \mathcal{K},f\in {{\mathcal{F}}_{2}}}}$, and preferences, ${{\left\{ P_{k,t-1}^{f} \right\}}_{k\in \mathcal{K},f\in {{\mathcal{F}}_{2}}}}$, in window $t-1$. Since each UDT data attribute has a temporal correlation, we utilize Autoencoders to process these data to reduce input dimension and abstract temporal features. The processed UDTs’ data are further input into the DDQN, consisting of one target and one main Q-network, for model training to obtain the appropriate clustering number ${{\Lambda}_{t}^{*}}$. Specifically, the main Q-network with network parameter $\theta $ takes the current processed UDTs’ data as input and outputs a set of Q-values. The target Q-network with network parameter $\theta '$ is a copy of the main Q-network that is periodically updated to match the parameters of the main Q-network. The target value, ${{y}_{t}}$, of DDQN is given by
\begin{equation}
	{{y}_{t}}={{r}_{t+1}}+\gamma Q\left( {{s}_{t+1}},\arg \max Q\left( {{s}_{t+1}},a;\theta  \right);\theta ' \right),
\end{equation}
where main Q-network $Q\left( s,a;\theta  \right)$ is utilized for the clustering number determination and target Q-network $Q\left( s,a;\theta ' \right)$ is used for the value evaluation. Here, $s$ and $a$ represent processed UDTs’ data and the clustering number, respectively. Since the objective of clustering is to reserve appropriate resources for different multicast groups to enhance system utility, we select the system utility that will be introduced in Section \ref{uti} as reward $r$. During the training process, the main Q-network is updated to minimize the difference between the predicted and true Q-values. The target Q-network is periodically updated to match the parameters of  main Q-network to help stabilize the training process. 

\subsubsection{User Clustering}
When the clustering number of multicast groups is determined, we can classify users into different multicast groups based on UDTs’ data similarities. We adopt the K-means++ algorithm to complete the user clustering process. Specifically, the Euclidean distance, $D\left( i,j \right)$, between UDTs $i$ and $j$ is first calculated as follows:
\begin{equation}\small
	D\left( i,j \right)={{\left\| h_{i}^{{}}-h_{j}^{{}} \right\|}_{2}}+{{\left\| Y_{i}^{{}}-Y_{j}^{{}} \right\|}_{2}}+{{\left\| w_{i}^{{}}-w_{j}^{{}} \right\|}_{2}}+{{\left\| P_{i}^{{}}-P_{j}^{{}} \right\|}_{2}}.
\end{equation}

Then, the Euclidean distance between the sampling UDT and the current clustering center is used to calculate the selection probability. Since the longer distance causes a higher selection probability as the new clustering center, the selection probability, $\Theta(i)$, of UDT $i$ is depicted as
\begin{equation}
	\Theta(i)=\frac{D{{\left( i,\tau \right)}^{2}}}{\sum\nolimits_{j}{D{{\left( j,\tau \right)}^{2}}}},
\end{equation}
where $\tau$ denotes the current clustering center. After several iterations, users with similar statuses are clustered into one multicast group. The set of constructed multicast group is denoted by ${{\Omega }_{t}}=\left\{ 1,\cdots ,\Lambda _{t}^{*} \right\}$\footnote{The computational complexity of the K-means++ algorithm is $O\left( {{\left( \Lambda _{t}^{*} \right)}^{2}}K+\pi \Lambda _{t}^{*}K \right)$, where $\pi$ and $K$ denote the number of iterations and users, respectively. The K-means++ algorithm has two benefits, i.e., low computational cost and fast convergence.}.

\subsubsection{Abstracted Information}
After constructing multicast groups, we need to abstract some useful information from each of them to predict resource demands in each resource reservation window. Recommended video lists and corresponding swipe probability distributions can well achieve this requirement. To obtain a good recommended video list, a good video recommendation mechanism not only needs to consider video popularity distributions but also users' preferences \cite{Gong}. The former can be directly obtained by analyzing view counts and engagement time from content providers. In window $t-1$, the popular video set is denoted by $\mathcal{V}=\left\{ 1,\cdots ,V \right\}$ and the corresponding popularity distribution is ${{\left\{ {{E}_{v}} \right\}}_{v\in \mathcal{V}}}$. The latter is time-series data and exists a relatively strong temporal correlation in each multicast group. The collected preference, $P_{k,t-1}^{f}$, in UDT $k$ is a $1\times C$ matrix, i.e., $\left\{ P_{k,t-1}^{f,1},\cdots ,P_{k,t-1}^{f,C} \right\}$, reflecting user $k$’s preferences of $C$ video types. To accurately estimate multicast group $g$’s preference matrix, ${{\left\{ \widehat{P}_{t,g}^{c} \right\}}_{c\in \mathcal{C}}}$, in current window $t$, we first calculate the users’ average preference on each kind of video in previous window $t-1$, and then integrate it with the discounted preference, $\widetilde{\lambda }\widehat{P}_{t-1,g}^{c}$, in previous window $t-1$, as follows:
\begin{equation}
\widehat{P}_{t,g}^{c}=\widetilde{\lambda }\widehat{P}_{t-1,g}^{c}+\frac{1}{{{K}_{g}}{{F}_{2}}}\sum\limits_{k=1}^{{{K}_{g}}}{\sum\limits_{f=1}^{{{F}_{2}}}{P_{k,t-1}^{f,c}}},\forall c\in \mathcal{C},
\end{equation}
where $\widetilde{\lambda}$ is a parameter ranging from 0 to 1. The recommended ranking, ${{\Re }_{t,g,v}}$, of video $v$ for multicast group $g$ in window $t$ combines the video popularity distribution and preference matrix, which is depicted as ${{\Re }_{t,g,v}}={{E}_{v}}\sum\nolimits_{c\in \mathcal{C}}{{{{I}}_{c}(v)}\widehat{P}_{t,g}^{c}}$. Here, ${{{I}}_{c}(v)}$ is an indicator function, representing if video $v$ belongs to type $c$, ${{{I}}_{c}(v)}=1$; otherwise, ${{{I}}_{c}(v)}=0$.

In each window, the recommended video list, $\tau_g$, for multicast group $g$ is constituted based on the recommended rankings in popular video set $\mathcal{V}$, and the corresponding list length is denoted by $\rho$. In addition, each video in the recommend list has a unique swipe probability distribution, $p_{g,t}^{v}\left( e \right)$, which can be abstracted by accumulating users’ swipe probability distributions in a multicast group, as follows: 
\begin{equation}
p_{g,t}^{v}\left( e \right)=\sum\limits_{k\in {{\mathcal{K}}_{g}}}{\sum\limits_{c\in \mathcal{C}}{{{{I}}_{c}(v)}p_{k,c}^{t}\left( e \right)}}.
\end{equation}


\subsection{UDT-Assisted Resource Demand Prediction}
Based on the abstracted grouping information, i.e., swipe probability distributions and recommended video lists, each multicast group’s bandwidth and computing resource demands can be predicted to help the network controller to facilitate an efficient resource reservation scheme.

\subsubsection{Bandwidth Resource Demand}
Since users’ swipe behaviors can cause part of videos not to be played and thus lead to a waste of bandwidth resources, we analyze users’ swipe probability distributions on the recommended video list to make an accurate bandwidth resource demand prediction.

First, the average engagement time, $W_{g}^{t}$, of multicast group $g$ at window $t$ is estimated based on the abstracted swipe probability distribution, which is expressed as
\begin{equation}
	W_{g}^{t}=\sum\nolimits_{v\in {{\tau }_{g}}}{\int_{0}^{\wp (v)}{\left( 1-p_{g,t}^{v}(e) \right)ede}}.
\end{equation}

Second, we need to predict the video traffic of each multicast group. Since one video can be transmitted to all users by multicast transmission in a multicast group, the video traffic of each user should not be accumulated. In addition, since the multicast video version adaptively changes with users’ dynamic channel conditions and buffer lengths, we select the average multicast video version, $\overline{l}$, in previous window $t-1$, to approximate the multicast video version in current window $t$. Based on the above analysis, the video traffic, $Y_{g}^{t}$, of multicast group $g$ at window $t$ is predicted by 
\begin{equation}
	Y_{g}^{t}=\sum\nolimits_{v\in {{\tau }_{g}}}{\int_{0}^{\wp (v)}{\left( 1-p_{g,t}^{v}(e) \right)\sum\nolimits_{l=1}^{\overline{l}}{z_{v}^{l}\left(e\right)}de}},
\end{equation}
where $z_{v}^{l}\left(e\right)$ is the file size of segment layer $l$ of video $v$ at video timestamp $e$. 

Based on the estimated average engagement time and video traffic, the bandwidth resource demand, $R_{g}^{t}$, is given by
\begin{equation}
	R_{g}^{t}=W_{g}^{t}/Y_{g}^{t}.
\end{equation}

\subsubsection{Computing Resource Demand}
Since the caching capacity of an edge server is limited, we only cache the basement layer of each recommended segment in the edge server to guarantee users’ basic watching requirements. Enhancement layers can be obtained by transcoding and then jointly transmitted with the basic layer to users to enhance the video quality. In this process, the computing consumption is predicted by
\begin{equation}
	Z_{g}^{t}=\mu \left( Y_{g}^{t}-\sum\nolimits_{v\in {{\tau }_{g}}}{\int_{0}^{\wp (v)}{\left( 1-p_{g,t}^{v}(e) \right)z_{v}^{0}\left( e \right)de}} \right),
\end{equation}
where $\mu $ is the computing density for video transcoding, and $z_{v}^{0}\left( e \right)$ is the file size of the basement layer of video $v$ at video timestamp $e$.

Based on the estimated average engagement time and computing consumption, the computing resource demand, $O_{g}^{t}$, is given by
\begin{equation}
	O_{g}^{t}=Z_{g}^{t}/Y_{g}^{t}.
\end{equation}

\section{User Satisfaction}
Since the system bandwidth and computing resources are limited, multicast groups’ resource demands may not be always satisfied. In addition, users usually have different sensitive degrees of rebuffering time and video quality \cite{Globe}, which indicates that the same reserved resources for various multicast groups can lead to different user satisfactions. Therefore, we analyze the relationship between reserved bandwidth and computing resources and user satisfaction for each multicast group, which can help the network controller to make a better resource reservation scheme to improve user satisfaction.

\begin{figure}
	\centering
	\subfigure[]{
		\includegraphics[width=0.23\textwidth]{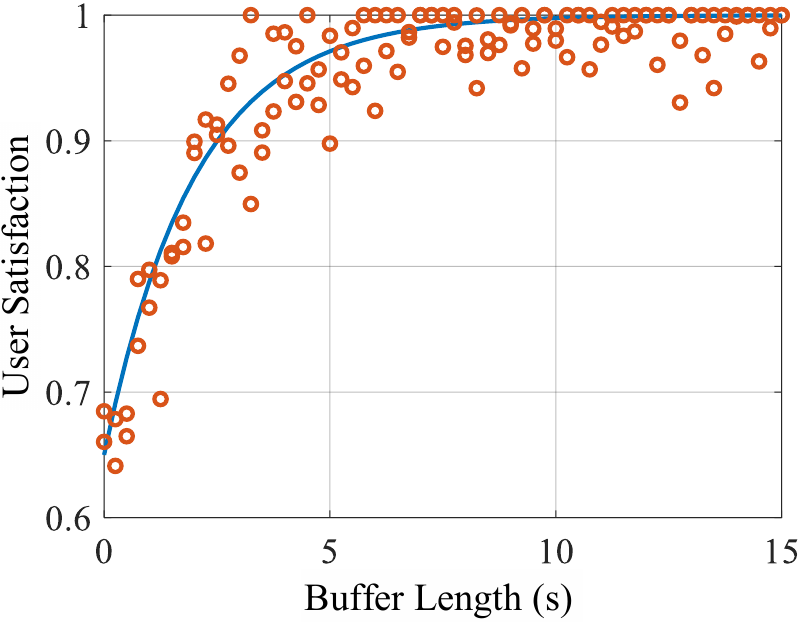}}
	\centering
	\subfigure[]{
		\includegraphics[width=0.23\textwidth]{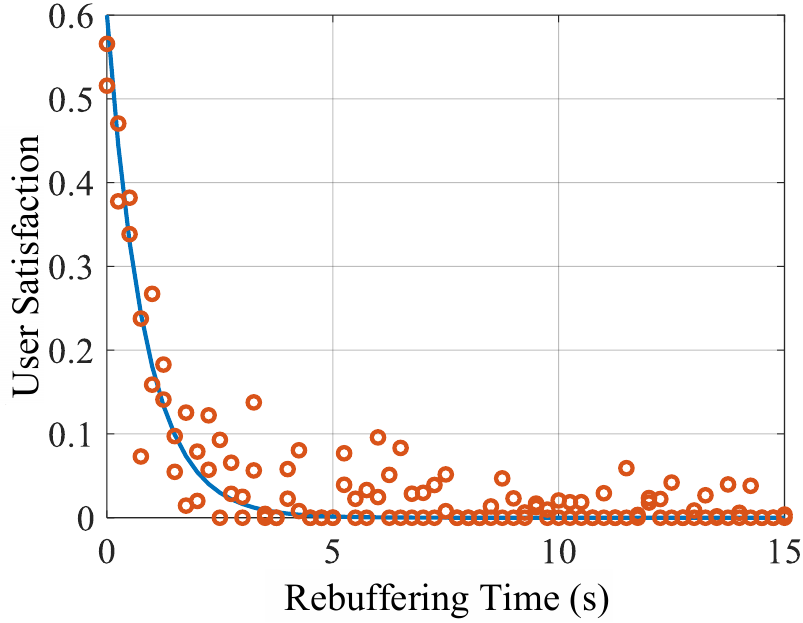}}
	\caption{User satisfaction over different buffer lengths and rebuffering time.}
	\label{fig:US}
\end{figure}

First, we analyze how reserved bandwidths affect user satisfaction based on observed users’ buffers. Specifically, if the downlink transmission rate estimated by reserved bandwidths is larger than the bandwidth resource demand, users’ buffer lengths can increase; otherwise, users’ buffers will gradually become empty and rebuffering time will increase. Therefore, the gap between the downlink transmission capability and the bandwidth resource demand is positively correlated with users' buffers. To measure how users’ buffers can affect user satisfaction, we select a part of users' watching and rating records from video dataset\footnote{Video quality assessment: https://live.ece.utexas.edu/research/ LIVEStallStudy/liveMobile.html}, and use the fitting function\footnote{Mathworks: https://www.mathworks.com/help/optim/ug/lsqcurvefit.html} to form a fitted curve from scattered points, as shown in Fig.~\ref{fig:US}.

It can be observed that user satisfaction is positively and negatively exponential with buffer length and rebuffering time, respectively. Therefore, we can conclude that user satisfaction is exponentially with the reserved bandwidths based on the intermediate variable, i.e., user's buffer. A parameterized sigmoid function is employed to characterize the exponential relationship between bandwidth reservation and user satisfaction as follows:
\begin{equation}\label{UB}
	U_{\text{B},g}^{t}\text{=}\frac{\text{1}}{\text{1+}\exp \left\{ -{{\xi }_{g}^{t}}\left({m}_{i}^{t}B\log \left( 1+\ell _{g}^{t} \right)-R_{g}^{t}\right) \right\}},
\end{equation}
where $\ell _{g}^{t}$ and $\xi _{g}^{t}$ are the average signal-to-interference plus noise ratio and the sensitivity of bandwidth reservation for multicast group $g$ in window $t$, respectively. Here, ${m}_{g}^{t}$ and $B$ are the reserved bandwidth number and size for multicast group $g$ in window $t$, respectively. When the downlink transmission rate of reserved bandwidths exceeds the bandwidth resource demand, user satisfaction will quickly increase and gradually tend to the highest value, and vice versa.

Second, we analyze how reserved VM instances affect user satisfaction. Specifically, if the transcoding capability estimated by reserved VM instances exceeds the computing resource demand, the video transcoding process can be quickly completed to guarantee that users can watch videos with satisfied video quality; otherwise, users will watch videos with low video quality and experience frequent quality variation. Since the video quality also has a similar exponential relationship with user satisfaction \cite{zhang2020satisfied}, we also adopt a parameterized sigmoid function to characterize the exponential relationship between VM instance reservation and user satisfaction as follows:
\begin{equation}
	U_{\text{V},g}^{t}=\frac{1}{1+\exp \left\{ -\vartheta _{g}^{t}\left({n}_{g}^{t}\omega-O_{g}^{t} \right) \right\}},
\end{equation}
where $\vartheta _{g}^{t}$ and ${n}_{g}^{t}$ represent the sensitivity of VM instance reservation and the number of reserved VM instances for multicast group $g$ in window $t$, respectively. Here, $\omega$ is the computing capacity of one VM instance.

\section{Problem Formulation}\label{prob}
\subsection{System Utility}\label{uti}
The system utility is defined to evaluate the performance of the UDT-assisted resource reservation scheme, including operation cost, reconfiguration cost, and user satisfaction.

\subsubsection{Operation cost}
The operation cost, $U_{\text{O}}^{t}$, refers to the cost of reserved bandwidths and VM instances for all multicast groups \cite{Wen} in window $t$, which can be expressed as
\begin{equation}
	U_{\text{O}}^{t}={{\varpi }_{1}}\sum\limits_{g=1}^{\Lambda_{t}^{*}}{m_{g}^{t}}+{{\varpi }_{2}}\sum\limits_{g=1}^{\Lambda_{t}^{*}}{n_{g}^{t}},
\end{equation}
where ${{\varpi }_{1}}$ and ${{\varpi }_{2}}$ represent the unit cost of using a bandwidth and a VM instance, respectively. 

\subsubsection{Reconfiguration Cost}
In each resource reservation window, if the previous resource reservation configuration for a multicast group cannot satisfy the current dramatically increasing resource demands, the network controller needs to configure bandwidths and VM instances for the multicast group, which incurs reconfiguration cost. The resource release cost is omitted since this process can be quickly completed with negligible cost. Based on the above analysis, the reconfiguration cost, $U_{\text{R}}^{t}$, can be expressed as
\begin{equation}\label{UR}
	U_{\text{R}}^{t}=\sum\limits_{g=1}^{\Lambda_{t}^{*}}{{{\varpi }_{3}}{{\left[ m_{g}^{t}-m_{g}^{t-1} \right]}^{+}}+{{\varpi }_{4}}{{\left[ n_{g}^{t}-n_{g}^{t-1} \right]}^{+}}}, 
\end{equation}
where ${{\varpi }_{3}}$ and ${{\varpi }_{4}}$ denote the unit cost of bandwidth and VM instance reconfigurations, respectively. Since the clustering number in the previous resource reservation window may be less than that in the current window, a part of elements in vectors ${{\mathcal{M}}^{t-1}}$ and ${{\mathcal{N}}^{t-1}}$ is denoted by 0 to guarantee dimension match, which can be expressed as
\begin{equation}
	m_{g}^{t-1}=0,n_{g}^{t-1}=0,\forall g\in \left[ \Lambda _{t-1}^{*}+1,\Lambda _{t}^{*} \right].
\end{equation}

The objective of system utility is to achieve low resource operation and reconfiguration costs and high user satisfaction. With Eq.~\eqref{UB}-\eqref{UR}, the system utility is defined as 
\begin{equation}
	{{U}^{t}}=\frac{1}{\Lambda _{t}^{*}}\sum\limits_{g=1}^{\Lambda _{t}^{*}}{\left( U_{\text{B},g}^{t}+{{\delta }_{1}}U_{\text{V},g}^{t} \right)}-{{\delta }_{2}}U_{\text{O}}^{t}-{{\delta }_{3}}U_{\text{R}}^{t},
\end{equation}
where ${{\delta }_{1}}$, ${{\delta }_{2}}$, and ${{\delta }_{3}}$ represent the weighted parameters to balance each system utility component.

\subsection{Problem Formulation}
Since multicast groups’ bandwidth and computing resource demands are dynamic due to users' diversified swipe behaviors and channel conditions, it is paramount to reserve the appropriate bandwidth and computing resources for each multicast group to guarantee a satisfied service quality. In each resource reservation window, our objective is to maximize the system utility within the limited bandwidth and computing resources, and the corresponding optimization problem is formulated as
\begin{align}\label{P0}
	{{\textbf{P}}_{\text{0}}:} \quad & \underset{{{\left\{ m_{g}^{t},n_{g}^{t} \right\}}_{g\in \Omega_t}}}{\mathop{\max }}\,{{U}^{t}} \\ 
	\text{s.t.} & \sum\nolimits_{g=1}^{\Lambda _{t}^{*}}{m_{g}^{t}}\le M,\forall t\in \mathcal{T}, \tag{18a} \label{1a} \\ 
	& \sum\nolimits_{g=1}^{\Lambda _{t}^{*}}{n_{g}^{t}}\le N,\forall t\in \mathcal{T}, \tag{18b} \label{1b}\\ 
	& m_{g}^{t}\in {{Z}^{+}},\forall g\in \Omega_t, t\in \mathcal{T}, \tag{18c} \label{1c} \\ 
	& n_{g}^{t}\in {{Z}^{+}},\forall g\in \Omega_t, t\in \mathcal{T}, \tag{18d} \label{1d} \\ 
	& \text{and}\; (16). \notag
\end{align}
Constraints~\eqref{1a} and \eqref{1b} are resource capacity constraints, which guarantee that the total reserved bandwidths and VM instances cannot exceed the system capacity.

\section{Proposed Resource Reservation Algorithm}
\subsection{Problem Transformation}
Problem ${\textbf{P}}_{\text{0}}$ is a nonlinear integer programming problem. To solve this problem, we first transform it into two subproblems. Specifically, since optimization variable $m_{g}^{t}$ is independent of optimization variable $n_{g}^{t}$, the problem can be decoupled into two independent minimization subproblems regarding bandwidth and VM instance reservation. The decoupled subproblems $\textbf{P}_{\text{1}}$ and $\textbf{P}_{\text{2}}$ are given by
\begin{align}
	\textbf{P}_{\text{1}}:&\underset{{{\left\{ m_{g}^{t} \right\}}_{g\in \Omega_{t}}}}{\mathop{\min }}\,\frac{1}{\Lambda _{t}^{*}}\sum\limits_{g \in \Omega_{t}} {\delta_3{\varpi }_{3}}{{\left[ m_{g}^{t}-m_{g}^{t-1} \right]}^{+}}+{\delta_2{\varpi }_{1}}{m_{g}^{t}}\notag\\&+{\frac{\text{1}}{-1-\exp \left\{ \xi _{g}^{t}\left( m_{g}^{t}B\log \left( 1+\ell _{g}^{t} \right)-R_{g}^{t} \right) \right\}}} \label{18}\\ 
	\text{s.t.} & \sum\limits_{g\in \Omega_t}{m_{g}^{t}}\le M,\forall t\in \mathcal{T}, \tag{19a} \\ 
	& m_{g}^{t}\in {{Z}^{+}},\forall g\in \Omega_t,t\in \mathcal{T}, \tag{19b} \\ 
	& m_{g}^{t-1}=0,\forall g\in \left[ \Lambda _{t-1}^{*}+1,\Lambda _{t}^{*} \right],t\in \mathcal{T}, \tag{19c}
\end{align}

\begin{align}
	\textbf{P}_{\text{2}}:&\underset{{{\left\{ n_{g}^{t} \right\}}_{g\in {{\Omega }_{t}}}}}{\mathop{\min }}\,\frac{1}{\Lambda _{t}^{*}}\sum\limits_{g\in {{\Omega }_{t}}}^{{}} {\delta_3{\varpi }_{4}}{{\left[ n_{g}^{t}-n_{g}^{t-1} \right]}^{+}}+{\delta_2{\varpi }_{2}}n_{g}^{t} \notag \\&+{\frac{\delta_1}{-1-\exp \left\{ \vartheta _{g}^{t}\left( n_{g}^{t}\omega-O_{g}^{t} \right) \right\}}} \label{19} \\ 
	\text{s.t.} & \sum\limits_{g\in {{\Omega }_{t}}}^{{}}{n_{g}^{t}}\le N,\forall t\in \mathcal{T}, \tag{20a} \\ 
	& n_{g}^{t}\in {{Z}^{+}},\forall g\in {{\Omega }_{t}},t\in \mathcal{T}, \tag{20b} \\ 
	& n_{g}^{t-1}=0,\forall g\in \left[ \Lambda _{t-1}^{*}+1,\Lambda _{t}^{*} \right],t\in \mathcal{T}. \tag{20c}
\end{align}

Then, we perform continuous processing on variables $m_g^t$ and $n_g^t$. If a continuous optimization problem is convex, the corresponding discrete optimization problem is also convex \cite{convex}. The objective function of subproblem $\textbf{P}_{\text{1}}$ consists of two parts, i.e., one related to the exponential term and one related to the approximately linear term. The former is expressed by $\Psi _{g}^{t}={-\text{1}}/({\text{1+}\exp \left\{ -\xi _{g}^{t}\left( m_{g}^{t}B\log \left( 1+\ell _{g}^{t} \right)-R_{g}^{t} \right) \right\}})$, and its convexity is related to the range of independent variable $m_g^t$. Specifically, the second derivative of $\Psi _{g}^{t}$ is shown in Eq.~\eqref{second_der}.

\begin{figure*}[hb]
	\noindent\rule{\textwidth}{0.4pt}  
	\centering
	\begin{equation}
		\frac{{{\partial }^{2}}\Psi _{g}^{t}}{{{\partial }^{2}}m_{g}^{t}} = \frac{\exp \left\{ -\xi _{g}^{t}\left( m_{g}^{t}B\log \left( 1+\ell _{g}^{t} \right)-R_{g}^{t} \right) \right\}{{\left( \xi _{g}^{t}B\log \left( 1+\ell _{g}^{t} \right) \right)}^{2}}}{{{\left( 1+\exp \left\{ -\xi _{g}^{t}\left( m_{g}^{t}B\log \left( 1+\ell _{g}^{t} \right)-R_{g}^{t} \right) \right\} \right)}^{3}}}\left( 1-\exp \left\{ -\xi _{g}^{t}\left( m_{g}^{t}B\log \left( 1+\ell _{g}^{t} \right)-R_{g}^{t} \right) \right\} \right).
		\label{second_der}
	\end{equation}
\end{figure*}
When $m_{g}^{t}\ge {R_{g}^{t}}/{B\log \left( 1+\ell _{g}^{t} \right)}$, we can have ${{{\partial }^{2}}\Psi _{g}^{t}}/{{{\partial }^{2}}m_{g}^{t}}\ge 0$ and function $\Psi _{g}^{t}$ is convex; otherwise, function $\Psi _{g}^{t}$ is concave. The latter is expressed by $\Gamma _{g}^{t}={{\delta }_{3}}{{\varpi }_{3}}{{\left[ m_{g}^{t}-m_{g}^{t-1} \right]}^{+}}+{{\delta }_{2}}{{\varpi }_{1}}m_{g}^{t}$, and it is convex due to the convexity of ${{\left[ \cdot  \right]}^{+}}$ function \cite{boyd}.

Since function $\Psi _{g}^{t}$ is not always convex in the whole range of independent variable $m_g^t$, we make an approximation to the concave part of  $\Psi _{g}^{t}$ to transform it into a convex function. Specifically, when $m_{g}^{t}<{R_{g}^{t}}/{B\log \left( 1+\ell _{g}^{t} \right)}$, we utilize a tangent to approximately substitute the concave part. The slope of tangent is the first derivative of  $\Psi _{g}^{t}$ when $m_{g}^{t}$ equals to ${R_{g}^{t}}/{B\log \left( 1+\ell _{g}^{t} \right)}\;$, which can be expressed by
\begin{equation}
	\kappa _{g}^{t}=-\frac{1}{4}\xi _{g}^{t}B\log \left( 1+\ell _{g}^{t} \right).
\end{equation}
The intersection of the tangent and the horizontal axis is given by
\begin{equation}
	b_{g}^{t}=\frac{\xi _{g}^{t}R_{g}^{t}-2}{\xi _{g}^{t}B\log \left( 1+\ell _{g}^{t} \right)}.
\end{equation}

To guarantee function $\Psi _{g}^{t}$ is always negative, independent variable $m_{g}^{t}$ needs to satisfy $m_{g}^{t}>b_{g}^{t}$. Based on this transformation, approximate function, $\widetilde{\Psi }_{g}^{t}$, can be expressed as a convex piecewise function, i.e.,
\begin{equation}
	\widetilde{\Psi }_{g}^{t}=\left\{ \begin{aligned}
		& \Psi_g^t,\; m_{g}^{t}\ge {R_{g}^{t}}/{{B\log \left( 1+\ell _{g}^{t} \right)}}, \\ 
		& \kappa _{g}^{t}m_{g}^{t}-0.5,\; b_{g}^{t}<m_{g}^{t}<{R_{g}^{t}}/{B\log \left( 1+\ell _{g}^{t} \right)}. \\ 
	\end{aligned} \right.
\end{equation}

Since the addition of convex functions $\widetilde{\Psi }_{g}^{t}$ and $\Gamma _{g}^{t}$ is still convex and constraints are linear, subproblem $\textbf{P}_{\text{1}}$ is transformed into a convex optimization subproblem $\textbf{P}_{\text{1}}'$, i.e.,
\begin{align}
	{{\mathbf{P}}_{\text{1}}'}: & \underset{{{\left\{ m_{g}^{t} \right\}}_{g\in {{\Omega }_{t}}}}}{\mathop{\min }}\,\frac{1}{\Lambda _{t}^{*}}\sum\limits_{g\in {{\Omega }_{t}}}{\widetilde{\Psi }_{g}^{t}+\Gamma _{g}^{t}}  \\
	\text{s.t.} & \quad m_{g}^{t}>b_{g}^{t}, \tag{25a} \\
	& \quad (19a),(19b),\text{and} \; (19c). \notag
\end{align}

Similarly, we can transform the exponential term, $\Xi _{g}^{t}$., in the objective function of subproblem $\textbf{P}_{\text{2}}$ into a convex piecewise function, which can be expressed as 
\begin{equation}
	\widetilde{\Xi }_{g}^{t}=\left\{ \begin{aligned}
		& \Xi_g^t, \; n_{g}^{t}\ge {O_{g}^{t}}/{\omega},\; \\ 
		& \widehat{\kappa }_{g}^{t}n_{g}^{t}-0.5, \; \widehat{b}_{g}^{t}<n_{g}^{t}<{O_{g}^{t}}/{\omega},\; \\ 
	\end{aligned} \right.
\end{equation}
where $\widehat{\kappa }_{g}^{t}=-\frac{1}{4}\vartheta _{g}^{t}\omega$ and $\widehat{b}_{g}^{t}=\frac{\vartheta _{g}^{t}O_{g}^{t}-2}{\vartheta _{g}^{t}\omega}$. The approximately linear term, $\widehat{\Gamma }_{g}^{t}$, in the objective function P2 is also convex due to the max function. Since the addition of convex functions $\widetilde{\Xi }_{g}^{t}$ and $\widehat{\Gamma }_{g}^{t}$ is still convex and constraints are linear, subproblem $\textbf{P}_{\text{2}}$ is transformed into a convex optimization subproblem $\textbf{P}_{\text{2}}'$, which can be expressed as
\begin{align}
	{{\mathbf{P}}_{\text{2}}'}: & \underset{{{\left\{ n_{g}^{t} \right\}}_{g\in {{\Omega }_{t}}}}}{\mathop{\min }}\,\frac{1}{\Lambda _{t}^{*}}\sum\limits_{g\in {{\Omega }_{t}}}{\widetilde{\Xi}_{g}^{t}+\widehat{\Gamma}_{g}^{t}}  \\
	\text{s.t.} & \quad n_{g}^{t}>\widehat{b}_{g}^{t},  \tag{27a} \\
	& \quad (20a),(20b),\text{and} \; (20c). \notag
\end{align}

Based on the linear approximation on the concave part,  the original optimization problem $\textbf{P}_{\text{0}}$ is transformed into a convex optimization problem.

\subsection{Fast Resource Reservation Algorithm}
Since the local optimal point is also the global optimal point for a convex optimization problem, we design a FS algorithm to find the local optimal bandwidth and VM instance reservation variables, i.e., ${{\left\{ m_{g}^{t} \right\}}_{g\in {{\Omega }_{t}}}}$ and ${{\left\{ n_{g}^{t} \right\}}_{g\in {{\Omega }_{t}}}}$. Specifically, each multicast group is first assigned $\left\lceil b_{g}^{t} \right\rceil $ bandwidths and $\left\lceil \widehat{b}_{g}^{t} \right\rceil$ VM instances. Then, in each iteration, the unassigned bandwidths and VM instances are sequentially assigned to the multicast group that can obtain the highest values of objective functions in subproblems $\textbf{P}_{\text{1}}'$ and $\textbf{P}_{\text{2}}'$, respectively. If the objective function value in the previous iteration is higher than that in the current iteration, the iteration process is terminated and local optimal resource reservation variables are the variables in the previous iteration. The specific algorithm is presented in Algorithm~\ref{alg1}.
\begin{algorithm}[t]
	\caption{Fast Scheduling (FS)}
	\label{alg1}
	\textbf{Initialize} the objective functions ${{U}^{t}}(\textbf{P}_{\text{1}}')$ and ${{U}^{t}}(\textbf{P}_{\text{2}}')$.
	
	\textbf{Input} ${{\Omega }_{t}}$, ${{\left\{ \ell _{g}^{t} \right\}}_{g\in {{\Omega }_{t}}}}$, ${{\left\{ R_{g}^{t} \right\}}_{g\in {{\Omega }_{t}}}}$, ${{\left\{ O_{g}^{t} \right\}}_{g\in {{\Omega }_{t}}}}$, $\left\lceil b_{g}^{t} \right\rceil$, $\left\lceil \widehat{b}_{g}^{t} \right\rceil$, $M$, $N$, $\Lambda _{t-1}^{*}$, and all weighted parameters.
	
	\textbf{Output} ${{\left\{ m_{g}^{t} \right\}}_{g\in {{\Omega }_{t}}}}$ and ${{\left\{ n_{g}^{t} \right\}}_{g\in {{\Omega }_{t}}}}$.
	
	\For{$g \in \Omega_{t}$}
	{
		Multicast group $g$ is assigned $\left\lceil b_{g}^{t} \right\rceil$ bandwidths and $\left\lceil \widehat{b}_{g}^{t} \right\rceil$ VM instances;
		
		Variables $m_g^t$ and $n_g^t$ take the values of $\left\lceil b_{g}^{t} \right\rceil$ and $\left\lceil \widehat{b}_{g}^{t} \right\rceil$, respectively;
	}
	
	Calculate the number of unassigned bandwidths and VM instances, i.e., $\widetilde{M} = M-\sum\nolimits_{g\in {{\Omega }_{t}}}{\left\lceil b_{g}^{t} \right\rceil}$ and $\widetilde{N} = N-\sum\nolimits_{g\in {{\Omega }_{t}}}{\left\lceil \widehat{b}_{g}^{t} \right\rceil}$;
	
	\For{$m=1:\widetilde{M}$}
	{
		\For{$g\in \Omega_t$}
		{
			Assign one bandwidth to multicast group $g$, and calculate ${{U}^{t}}(\textbf{P}_{\text{1},g}')$;
		}
		Update variable $m_g^t$ by $m_g^t = m_g^t + 1$ with the minimum value of the objective function, i.e., ${{g}^{*}}=\arg \underset{g}{\mathop{\min }}\,{{U}^{t}}(\textbf{P}_{\text{1},g}')$; 
		
		\If{${{U}^{t}}(\textbf{\rm{P}}_{\text{\rm{1}}}')^{(m)} \ge {{U}^{t}}(\textbf{\rm{P}}_{\text{\rm{1}}}')^{(m-1)}$ }
		{
			Stop the iteration and return ${{\left\{ m_{g}^{t} \right\}}_{g\in {{\Omega }_{t}}}}$;
		}
	}
	
		\For{$n=1:\widetilde{N}$}
	{
		\For{$g\in \Omega_t$}
		{
			Assign one VM instance to multicast group $g$, and calculate ${{U}^{t}}(\textbf{P}_{\text{2},g}')$;
		}
		Update variable $n_g^t$ by $n_g^t = n_g^t + 1$ with the maximum value of the objective function, i.e., ${{g}^{*}}=\arg \underset{g}{\mathop{\max }}\,{{U}^{t}}(\textbf{P}_{\text{2}, g}')$; 
		
		\If{${{U}^{t}}(\textbf{\rm{P}}_{\text{\rm{2}}}')^{(n)} \ge {{U}^{t}}(\textbf{\rm{P}}_{\text{\rm{2}}}')^{(n-1)}$ }
		{
			Stop the iteration and return ${{\left\{ n_{g}^{t} \right\}}_{g\in {{\Omega }_{t}}}}$;
		}
	}
\end{algorithm}

\subsection{Computational Complexity Analysis}
The proposed FS algorithm needs to find the local optimal points for bandwidth and VM instance reservation. The analysis of computational complexity is as follows. First, the computational complexity of initial resource reservation for each multicast group is $O\left( \Lambda _{t}^{*} \right)$. Then, the computational complexity of assigning the rest bandwidths is $O\left( \widetilde{m}\Lambda _{t}^{*} \right)$, where $\widetilde{m}$ is a positive value less than $\widetilde{M}$ since the bandwidth assignment can be momentarily terminated before all bandwidths are completely assigned. Next, the computational complexity of assigning the rest VM instances is $O\left( \widetilde{n}\Lambda _{t}^{*} \right)$, where $\widetilde{n}$ is a positive value less than $\widetilde{N}$ since the VM instance assignment can be momentarily terminated before all VM instances are completely assigned. Finally, the overall computational complexity of FS algorithm is $O\left( \Lambda _{t}^{*}+\widetilde{m}\Lambda _{t}^{*}+\widetilde{n}\Lambda _{t}^{*} \right)$.

\section{Simulation Results}
\subsection{Simulation Setup}
We conduct extensive simulations on the real-world dataset to evaluate the performance of the proposed UDT-assisted resource reservation scheme. The main simulation parameters are presented in Table \ref{sim}. 
\begin{table}[t]
	\centering
	\setlength{\tabcolsep}{4.5pt}
	\caption{Simulation Parameters}
	\label{sim}
	\begin{tabular}{c c|c c |c c}
		\hline
		\hline
		\textbf{Parameter}                 & \textbf{Value} & \textbf{Parameter}                 & \textbf{Value} & \textbf{Parameter}        & \textbf{Value}  \\ \hline
		$M$ &	15	  & $T$	& 5 min	& ${{\varpi }_{1}}$	& 0.5 \\ 
		$B$	& 2 MHz	  & $\rho $ 	& 50	& ${{\varpi }_{2}}$	& 0.5 \\
		$N$	& 10	  &$C$	& 8	& ${{\varpi }_{3}}$	& 0.7 \\
		$\omega$	& 2 G Cycle/s	& $F_1$	& 150	& ${{\varpi }_{4}}$	& 1 \\
		$K$	& 60	& $F_2$	& 5	& ${{\delta }_{1}}$	& 1.5  \\
		$V$	& 1000	& $\lambda$	& 0.4	& ${{\delta }_{2}}$	& 0.3  \\
		$\mu$ &	2 G Cycle/Mb	& $\widehat{\lambda}$	& 0.3	& ${{\delta }_{3}}$	& 0.3 \\
				\hline
	\end{tabular}
\end{table}
The key components of the simulation are introduced as follows.

\begin{figure}[t]
	\centering
	\includegraphics[width=8.8cm]{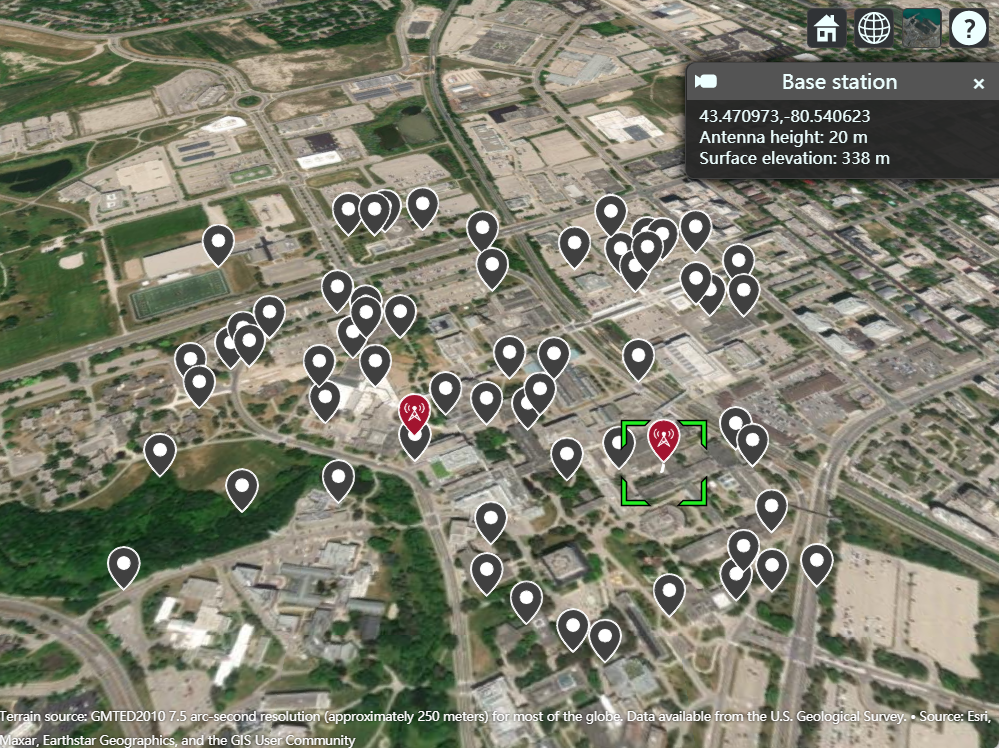}
	\caption{The simulation scene, where BSs and users are represented by red and gray icons, respectively.}
	\label{fig:BSs}
\end{figure}

We adopt the short video streaming dataset\footnote{ACM MM Grand Challenges: https://github.com/AItransCompetition/Short-Video-Streaming-Challenge/tree/main/data} to obtain users’ swipe behaviors and the user satisfaction dataset\footnotemark[1] to fit the user satisfaction function. We sample 1000 short videos from the YouTube 8M dataset\footnote{YouTube 8M dataset: https://research.google.com/youtube8m/index.html}, which includes 8 video types, i.e., Entertainment, Games, Food, Sports, Science, Dance, Travel, and News. Each video has a duration of 15~$sec$ and is encoded into four versions by the H. 265 encoder. We consider the scenario where two BSs are deployed at the University of Waterloo (UW) campus and users’ initial positions are randomly and uniformly generated around two BSs, as shown in Fig.~\ref{fig:BSs}. Each user moves along a prescribed path within the UW campus at a speed of 2$\sim$5 $km/h$, and the corresponding channel path loss is obtained by the propagationModel at Matlab. The transmission power and noise power are set to 27 dBm and -174 dBm, respectively.

Since the dimensions of different UDT data attributes are different, we construct four Autoencoder models for data compression. Take the Autoencoder model for the compression of locations as an example, the encoder is composed of two Conv2D layers, each using a ReLU activation function and a 'same' padding. The first Conv2D layer has 32 filters with a kernel size of (1, 3), while the second Conv2D layer has 64 filters with the same kernel size. After the Conv2D layers, a Flatten layer is applied to transform the multi-dimensional tensor into a one-dimensional tensor. The flattened tensor is then passed through a Dense layer with 60 neurons and a linear activation function, resulting in a compressed representation of input data in the latent space. The decoder is designed to reconstruct input data from the latent space representation. It starts with a Dense layer having $60 \times150 \times64$ neurons, followed by a Reshape layer to convert the tensor back to the shape (60, 150, 64). Then, a Conv2DTranspose layer with 32 filters, kernel size (1, 3), ReLU activation function, and 'same' padding is applied. Finally, the output layer is another Conv2DTranspose layer with 2 channels (matching the input data), kernel size (1, 3), linear activation function, and 'same' padding. The compressed UDT data is then input to DDQN to determine the clustering number. The detailed Autoencoder and DDQN parameters are shown in Table \ref{Auto} and \ref{DDQN}, respectively. 

We compare the proposed UDT-assisted resource reservation scheme with the following benchmark schemes: (1) without DT (WDT), where multicast groups are constructed and updated based on users’ preferences and locations, and bandwidth and VM instance reservation is based on historical video requests without considering users' swipe behaviors; (2) density-based spatial clustering of applications with noise \cite{DBSCAN} and FS algorithm (DBSCAN-FS); (3) DT-based user clustering scheme, and branch- and bound-based \cite{BBM} scheduling algorithm (DT-BBS). Since the branch- and bound- method can obtain the optimal solution to 0-1 integer programming problem, the DT-BBS also represents our performance upper bound but the disadvantage is time-consuming; (4) DT-based user clustering scheme, and branching dueling Q-network-based scheduling algorithm (DT-BDQN), which can solve the high-dimensional resource scheduling problem by splitting the action space \cite{BDQ}.

\begin{table}[t]
		\centering
	\setlength{\tabcolsep}{3.5pt}
	\caption{Autoencoder parameters}
	\label{Auto}
	\begin{tabular}{c|c|c|c|c}
		\hline
		\hline
		\textbf{Model}                                                                                                  & \textbf{Layer name}      & \textbf{NN units}                                                                                                   & \textbf{Activation} & \textbf{Padding} \\ \hline
		\multirow{6}{*}{\begin{tabular}[c]{@{}c@{}}Compression \\ for channel \\ conditions, \\ swipe \\ timestamps, \\ and prefers \end{tabular}} & Conv1D          & 32, 3                                                                                                      & ReLu       & same    \\ \cline{2-5} 
		& Conv1D          & 64, 3                                                                                                      & ReLu       & same    \\ \cline{2-5} 
		& Dense           & 60                                                                                                         & Linear     & /       \\ \cline{2-5} 
		& Dense           & $60\times64$                                                                                                       & ReLu       & /       \\ \cline{2-5} 
		& Conv1DTranspose & 32, 3                                                                                                      & ReLu       & same    \\ \cline{2-5} 
		& Conv1DTranspose & \begin{tabular}[c]{@{}c@{}}150/9e3/160, 3\end{tabular} & Linear     & same    \\ \hline
		\multirow{6}{*}{\begin{tabular}[c]{@{}c@{}}Compression \\ for locations\end{tabular}}                                             & Conv1D          & 32, $1\times3$                                                                                                 & ReLu       & same    \\ \cline{2-5} 
		& Conv1D          & 64, $1\times3$                                                                                                    & ReLu       & same    \\ \cline{2-5} 
		& Dense           & 60                                                                                                         & Linear     & /       \\ \cline{2-5} 
		& Dense           & 5.76e5                                                                                                 & ReLu       & /       \\ \cline{2-5} 
		& Conv1DTranspose & 32, $1\times3$                                                                                                    & ReLu       & same    \\ \cline{2-5} 
		& Conv1DTranspose & 2, $1\times3$                                                                                                    & Linear     & same    \\ \hline
	\end{tabular}
\end{table}

\begin{table}[t]
	\centering
	\setlength{\tabcolsep}{4.5pt}
	\caption{DDQN Parameters}
	\label{DDQN}
	\begin{tabular}{c c | c c}
		\hline
		\hline
		\textbf{Parameter}                 & \textbf{Value} &\textbf{Parameter}                 & \textbf{Value}  \\ \hline
		Memory size &	2000 &	Initial exploration rate	& 1   \\
		Discount rate &	0.95 &	Exploration decay rate &	0.995 \\
		Episode length &	90 &	NN layer connection &	FC\\
		Number of Episodes &	300 &	Number of hidden layers &	3 \\
		Learning rate &	0.001 &	Activation function &	ReLU \\
		Mini-batch size &	32 &	Number of neurons & 
		\begin{tabular}[c]{@{}l@{}}512$\times$256$\times$ \\ 128$\times$64$\times$10 \end{tabular} \\
		\hline
	\end{tabular}
\end{table}

\subsection{Clustering Performance Evaluation}

\begin{figure}[t]
	\centering
	\includegraphics[width=6.5cm]{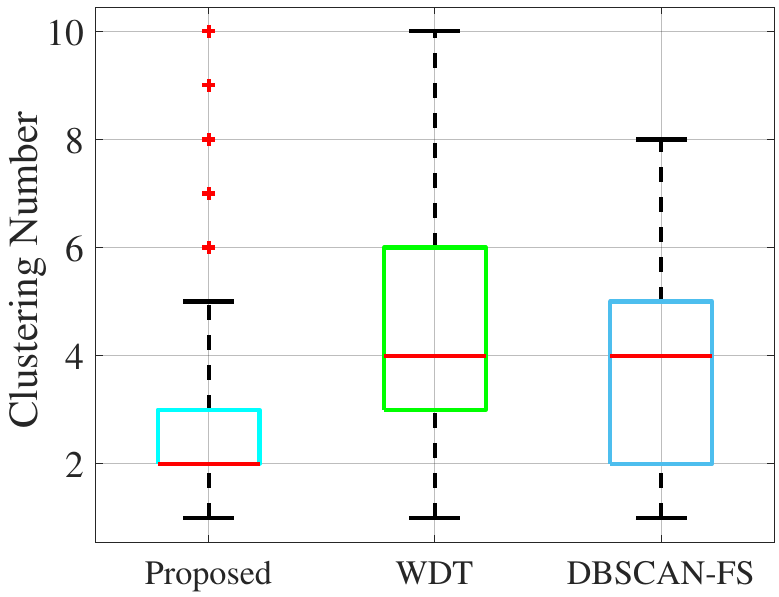}
	\caption{Clustering number comparison.}
	\label{fig:clunum}
\end{figure} 
In this section, we will compare the clustering number, the user density in each multicast group, and the convergence performance of proposed clustering algorithm, respectively.

As shown in Fig. \ref{fig:clunum}, we present the clustering number distribution in the boxplot for 90 resource reservation windows. It can be observed that the proposed scheme can achieve lower first-quartile, median, third-quartile, and maximum values compared with other schemes. Our proposed scheme demonstrates superior performance with relatively minimal variations, while the WDT scheme exhibits a larger fluctuation. This can be attributed to the unique capability of UDTs to effectively extract users’ swipe probability distributions. Consequently, our proposed scheme empowered by UDTs can well adapt to changes in swipe behaviors and network conditions. Overall, the proposed scheme offers a more robust and efficient solution for managing multicast groups in the face of dynamic changes in user statuses.

\begin{figure}[t]
	\centering
	\includegraphics[width=7.2cm]{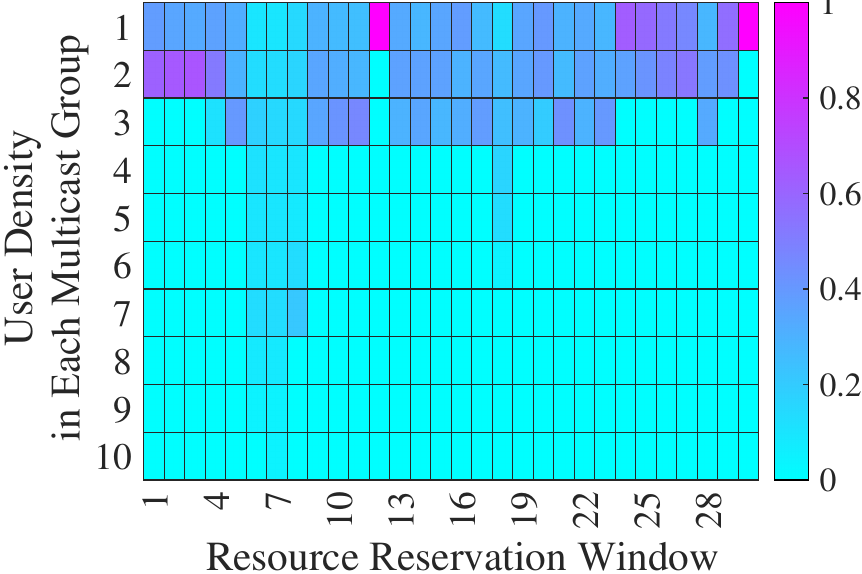}
	\caption{User density in each multicast group.}
	\label{fig:density}
\end{figure} 

As illustrated in Fig. \ref{fig:density}, we present the user density of proposed UDT-assisted clustering scheme during each resource reservation window. The darker the color, the higher the user density, and vice versa. Overall, the variation trend of user density in each multicast group is relatively gradual. This can be attributed to two primary reasons. First, reconfiguring bandwidth and computing resources for a multicast group entails additional network overheads. Second, to accommodate diverse user demands and preferences, the number of users within a multicast group is inherently limited. Furthermore, we observe that certain multicast groups exhibit significantly higher user density than others. This phenomenon arises because the majority of users own similar characteristics. By clustering the similar users for multicast transmission and video transcoding, network traffic burden can be effectively relieved.

\begin{figure}[t]
	\centering
	\includegraphics[width=6.8cm]{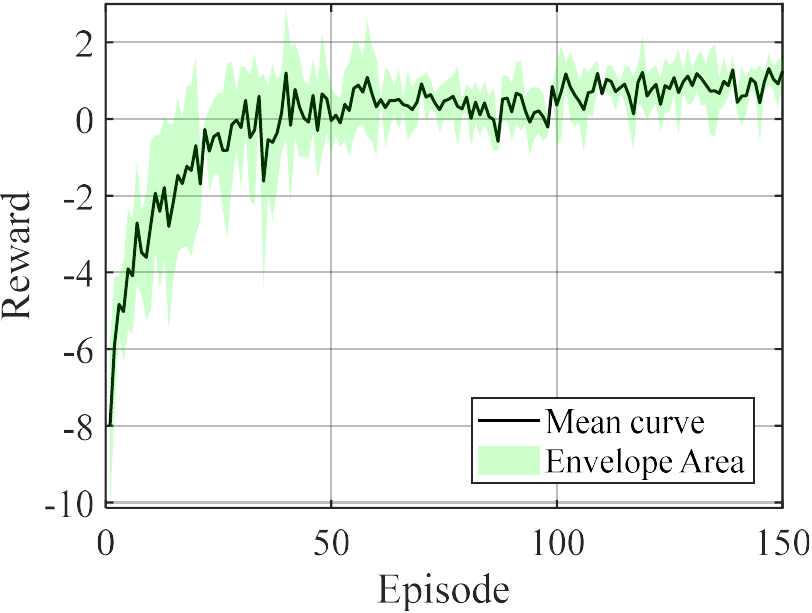}
	\caption{Convergence performance of DDQN-based clustering algorithm.}
	\label{fig:reward}
\end{figure} 
As shown in Fig. \ref{fig:reward}, we present the convergence curve of the DDQN-based clustering algorithm for determining the number of multicast groups. We conducted three trials of training to draw the corresponding envelope area and mean curve. Each episode consists of 90 steps, and the corresponding reward is the average reward for all steps within an episode. It can be observed that as the number of episodes increases, the reward gradually grows larger. When the number of episodes approaches nearly 70, the reward converges to a stable state, indicating that the DDQN-based clustering algorithm can effectively extract user similarity from user statuses to determine the number of multicast groups.

\subsection{Abstracted Swipe Probability Distribution}

\begin{figure*}[htbp!]
	\centering
	\subfigure[Resource Reservation Window 1]{
		\includegraphics[width=0.32\textwidth]{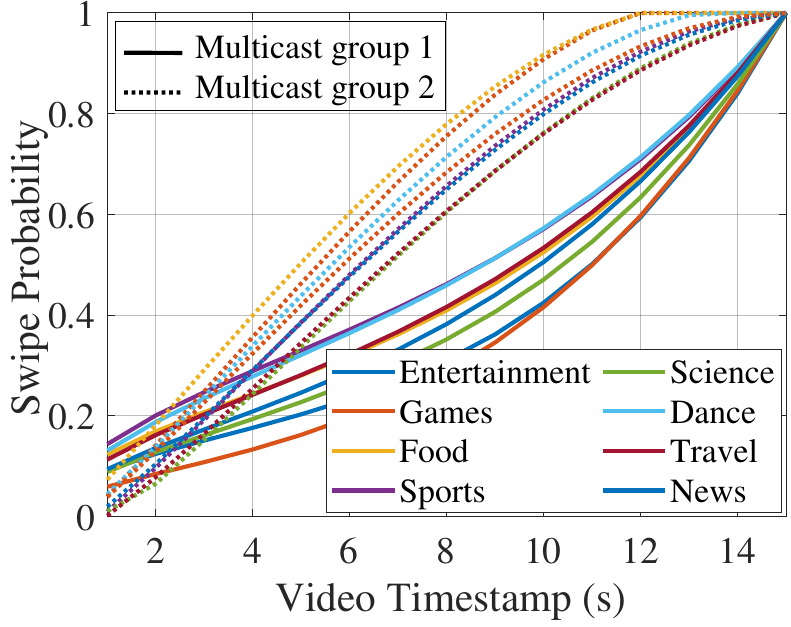}}
	\centering
	\subfigure[Resource Reservation Window 5]{
		\includegraphics[width=0.32\textwidth]{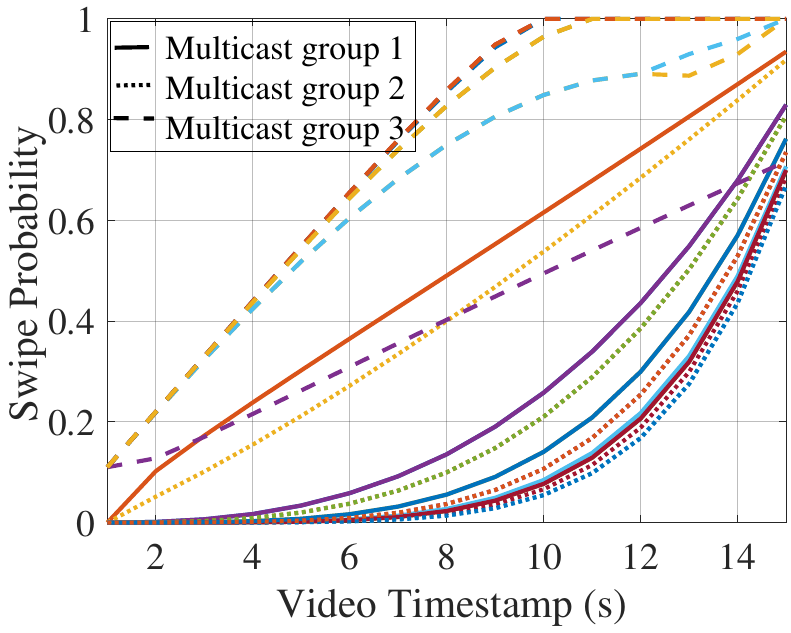}}
	\centering
	\subfigure[Resource Reservation Window 9]{
		\includegraphics[width=0.32\textwidth]{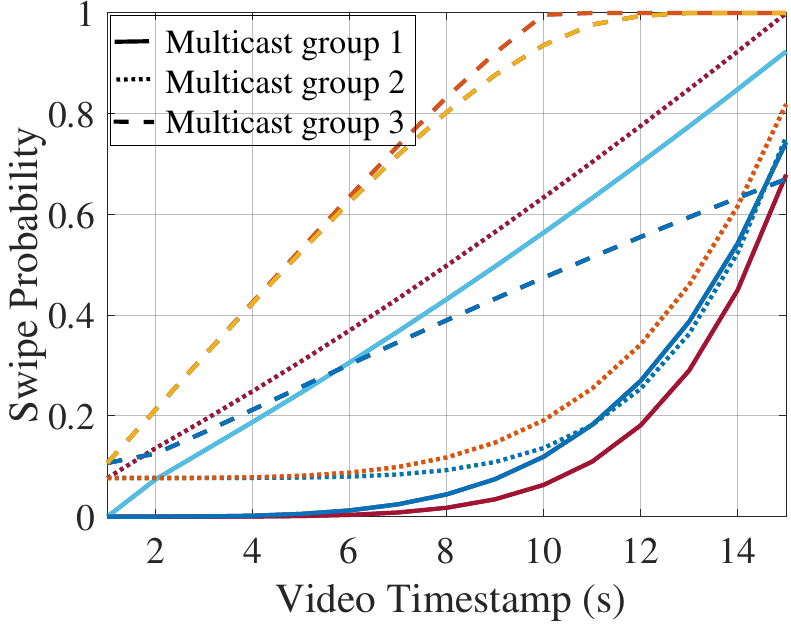}}
	\caption{Swipe probability distribution abstracted by DT.}
	\label{fig:swipe}
\end{figure*}

As illustrated in Fig. \ref{fig:swipe}, we present the swipe probability distribution extracted via UDTs across various resource reservation windows, where different line styles denote different multicast groups. Leveraging users’ status information stored in UDTs, we can infer their swipe probability distributions and cluster them with similar statuses into the same multicast group. Fig.~\ref{fig:swipe}(a) reveals that users in multicast group 1 demonstrate similar swipe probability distributions as those in multicast group 2 during the initial phase, but a noticeable divergence ensues over time. This suggests that UDTs can effectively differentiate swipe behaviors of various users from a global perspective, thereby facilitating more precise information provision for resource reservation. Fig.~\ref{fig:swipe}(b) and (c) exhibit the swipe probability distributions among three multicast groups. An overlap of swipe probabilities on certain types of videos occurs, indicating that over time, user behavior with respect to swipe probabilities for different types of videos gradually converges. This also implies a diminishing influence of video type on the swipe probability. Such information can be effectively captured by UDTs, providing valuable input to the network controller for more accurate resource reservation.

\subsection{System Utility Performance Evaluation}

\begin{figure*}[htbp]
	\centering
	\subfigure[Bandwidth Reservation Satisfaction (BRS)]{
		\includegraphics[width=0.32\textwidth]{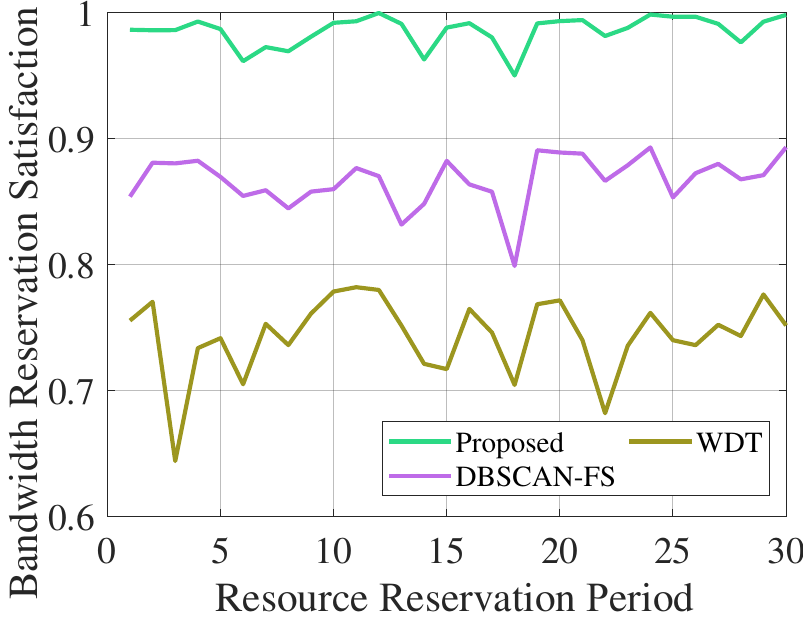}}
	\centering
	\subfigure[VM Reservation Satisfaction (VMRS)]{
		\includegraphics[width=0.32\textwidth]{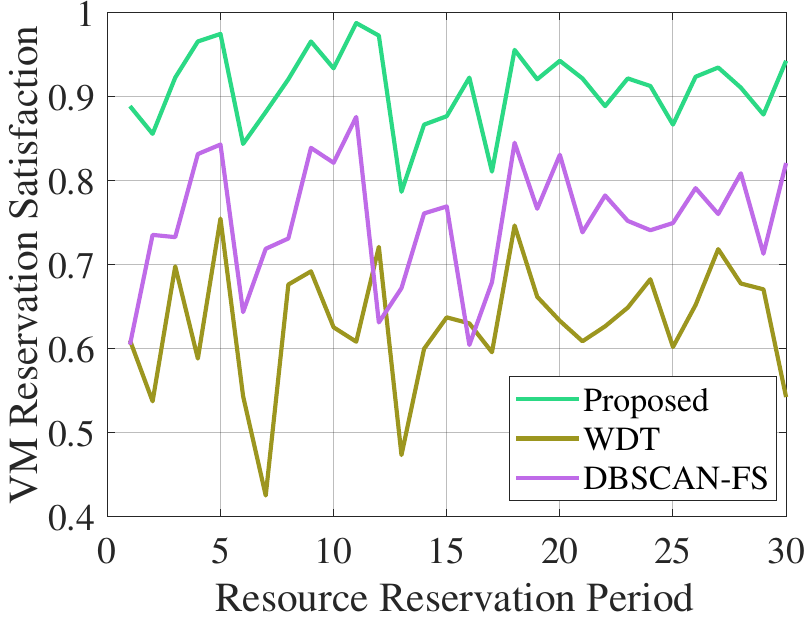}}
	\centering
	\subfigure[Bandwidth Operation Cost (BOC)]{
		\includegraphics[width=0.32\textwidth]{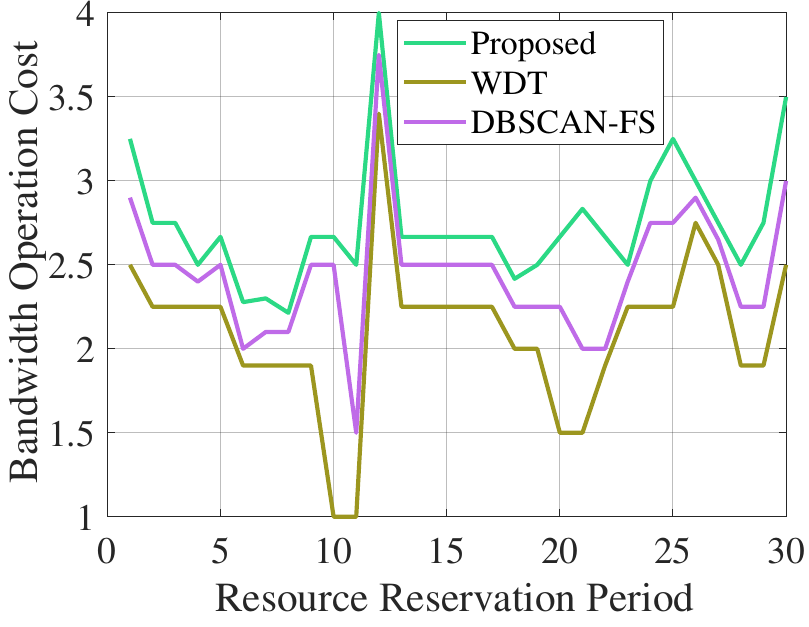}}
	\centering
	\subfigure[VM Operation Cost (VMOC)]{
		\includegraphics[width=0.32\textwidth]{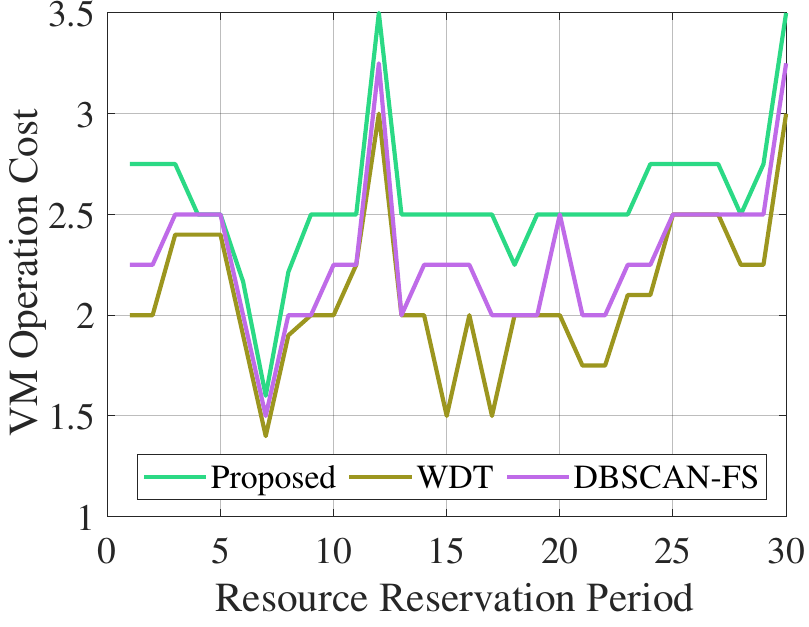}}
	\centering
	\subfigure[Bandwidth Reconfiguration Cost (BRC)]{
		\includegraphics[width=0.32\textwidth]{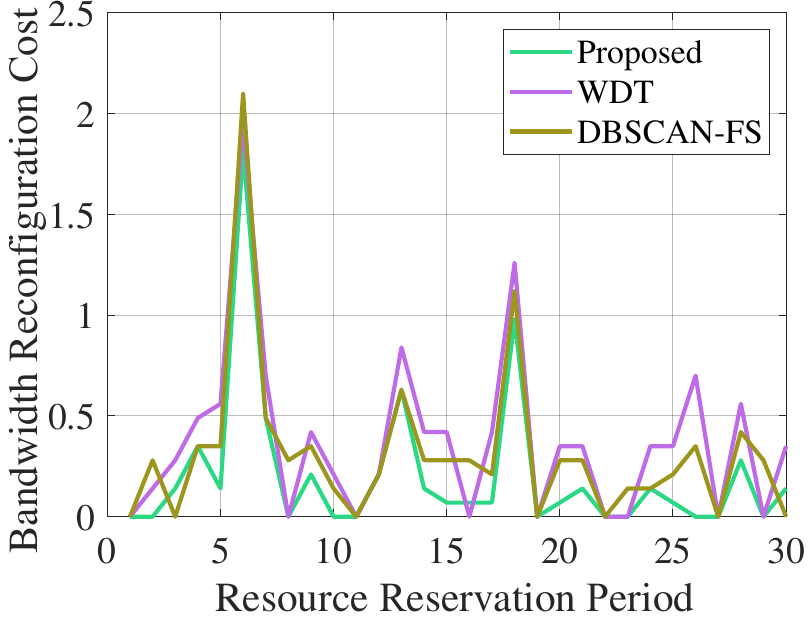}}
	\centering
	\subfigure[VM Reconfiguration Cost (VMRC)]{
		\includegraphics[width=0.32\textwidth]{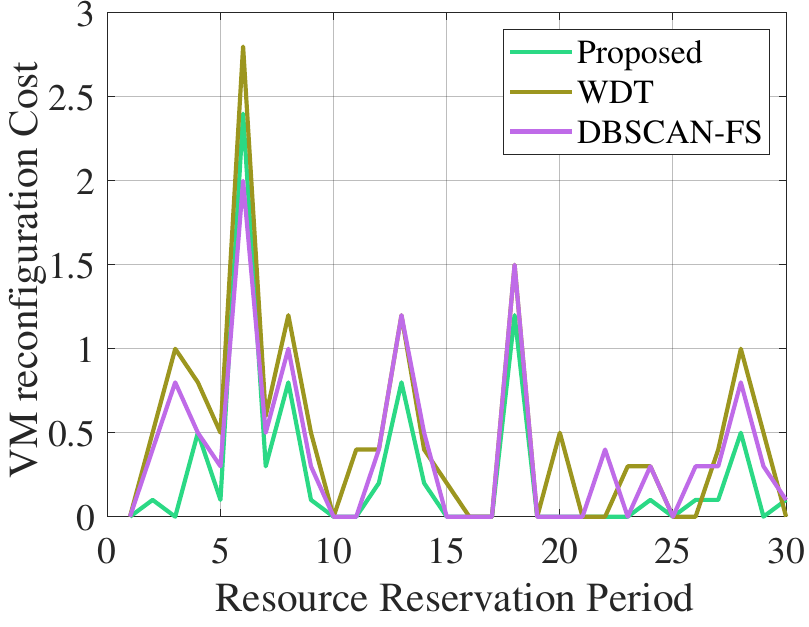}}
	\caption{The performance comparison of each component in the system utility.}
	\label{fig:metrics}
\end{figure*}

As illustrated in Fig. \ref{fig:metrics}, we first present the performance comparison of various system utility components under different schemes. With respect to bandwidth and VM reservation satisfaction, the proposed scheme can achieve a relatively high satisfaction level. Especially, BRS exhibits less fluctuation compared with VMRS. This is because the number of reserved VMs directly affects the speed of video transcoding, leading to variations in video quality and buffer length. Consequently, when the number of reserved VMs is insufficient, user satisfaction tends to fluctuate more significantly. In comparison to other schemes, WDT has the lowest satisfaction level. This is attributed to the lack of UDTs, which disables the network controller from swiftly and accurately analyzing users’ similarities from their historical data to precisely construct multicast groups. As a result, bandwidth and VM instance reservation cannot meet the actual resource demands of each multicast group, thereby leading to lower satisfaction. 

In terms of bandwidth and VM operation cost, the proposed method is relatively high. Firstly, to ensure the satisfaction level of resource reservation, the network controller needs to reserve more bandwidth and VMs. Secondly, to avoid the frequent reconfiguration cost, the network controller tends to reserve a relatively large amount of bandwidth and VMs. As for reconfiguration cost, the proposed scheme is relatively low. This is because DT can globally analyze users’ similarities and swipe behaviors, and group users on a larger time scale, thereby effectively avoiding frequent resource reconfigurations. 

\begin{figure}[t]
	\centering
	\includegraphics[width=6.8cm]{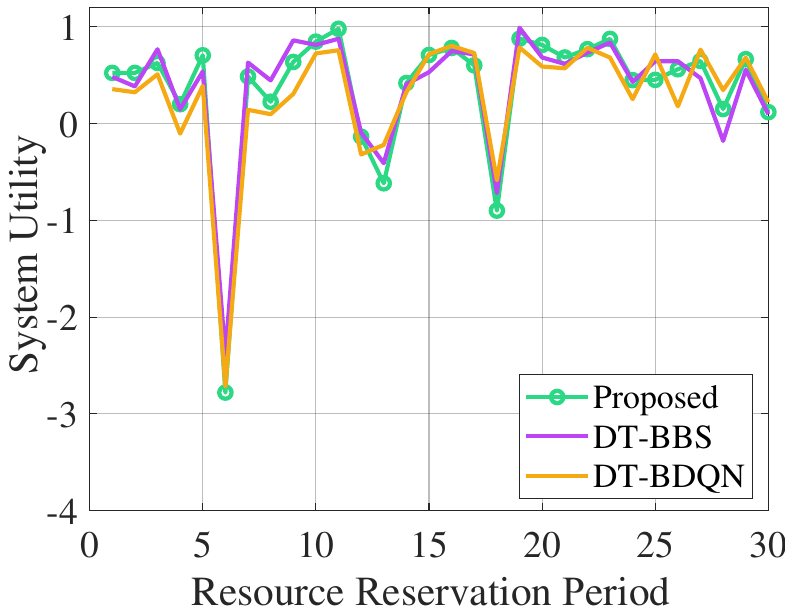}
	\caption{System utility comparison.}
	\label{fig:utility}
\end{figure} 

Then, we compare the system utility corresponding to different schemes in Fig. \ref{fig:utility}. It is observed that the proposed scheme can achieve quite a high system utility, although it remains generally lower than the DT-BBS scheme. This is because the DT-BBS scheme employs the branch and bound method to solve the 0-1 integer programming problem of bandwidth and VM reservation, which yields an optimal solution. Therefore, the corresponding system utility is generally the highest. The DT-BDQN scheme can also achieve a very close system utility compared with other schemes, indicating the BDQN algorithm can well learn resource scheduling strategies adapting to the dynamics of network conditions and user behaviors.

\begin{table}[t]
	\setlength{\tabcolsep}{4.0pt}
	\caption{Performance metric comparison}
	\label{metric}
	\begin{tabular}{c|cl|cl|cl|cl|cl}
		\hline
		\hline
		\textbf{}                                                         & \multicolumn{2}{c|}{Proposed}      & \multicolumn{2}{c|}{WDT}           & \multicolumn{2}{c|}{DBSCAN-FS}     & \multicolumn{2}{c|}{DT-BBS}        & \multicolumn{2}{c}{DT-BDQN}      \\ \hline
		BRS                                                               & \multicolumn{2}{c|}{\textbf{0.99}} & \multicolumn{2}{c|}{0.74}          & \multicolumn{2}{c|}{0.87}          & \multicolumn{2}{c|}{\textbf{0.99}} & \multicolumn{2}{c}{0.98}         \\ \hline
		VMRS                                                              & \multicolumn{2}{c|}{0.91}          & \multicolumn{2}{c|}{\textbf{0.63}} & \multicolumn{2}{c|}{0.75}          & \multicolumn{2}{c|}{\textbf{0.93}} & \multicolumn{2}{c}{0.91}         \\ \hline
		BOC                                                               & \multicolumn{2}{c|}{2.74}          & \multicolumn{2}{c|}{\textbf{2.1}}  & \multicolumn{2}{c|}{2.45}          & \multicolumn{2}{c|}{2.85}          & \multicolumn{2}{c}{2.8}          \\ \hline
		VMOC                                                              & \multicolumn{2}{c|}{\textbf{2.57}} & \multicolumn{2}{c|}{\textbf{2.11}} & \multicolumn{2}{c|}{2.29}          & \multicolumn{2}{c|}{\textbf{2.59}} & \multicolumn{2}{c}{2.63}         \\ \hline
		BRC                                                               & \multicolumn{2}{c|}{0.21}          & \multicolumn{2}{c|}{0.38}          & \multicolumn{2}{c|}{\textbf{0.32}} & \multicolumn{2}{c|}{0.22}          & \multicolumn{2}{c}{\textbf{0.2}} \\ \hline
		VMRC                                                              & \multicolumn{2}{c|}{0.25}          & \multicolumn{2}{c|}{0.5}           & \multicolumn{2}{c|}{0.4}           & \multicolumn{2}{c|}{\textbf{0.23}} & \multicolumn{2}{c}{0.27}         \\ \hline
		\begin{tabular}[c]{@{}c@{}}Average \\ system utility\end{tabular} & \multicolumn{2}{c|}{0.36}          & \multicolumn{2}{c|}{-0.34}         & \multicolumn{2}{c|}{-0.04}         & \multicolumn{2}{c|}{\textbf{0.37}} & \multicolumn{2}{c}{0.29}         \\ \hline
		\begin{tabular}[c]{@{}c@{}}Average \\ runtime (s)\end{tabular}    & \multicolumn{2}{c|}{1.54}          & \multicolumn{2}{c|}{0.74}          & \multicolumn{2}{c|}{\textbf{1.15}} & \multicolumn{2}{c|}{9.63}          & \multicolumn{2}{c}{2.12}         \\ \hline
	\end{tabular}
\end{table}

Finally, we compare the system performance metrics under different schemes, including average BRS, VMRS, BOC, VMOC, BRC, VMRC, system utility, and runtime as shown in Table \ref{metric}. The best performance corresponding to each metric is highlighted in bold. As can be observed, although our proposed scheme cannot achieve the best performance in all metrics, it ensures a high average BRS, VMRS, and system utility. Especially, the system utility is very close to the DT-BBS scheme, significantly higher than that of the WDT and DBSCAN-FS schemes, but the required system runtime is much lower than that of the DT-BBS scheme. This demonstrates that our proposed scheme can quickly adapt to the dynamics of the network conditions and users’ swipe behaviors to make timely adjustments to resource reservation.

\section{Conclusion}
In this paper, we have studied a novel resource management issue to enhance the MSVS performance. We have proposed a UDT-assisted resource reservation scheme to abstract the swipe probability distribution and recommended video list for the bandwidth and computing resource demand prediction. Furthermore, we have proposed a user satisfaction model by taking the user’s personalized preference and service sensitivity into account. A low-complexity resource scheduling algorithm has been designed to determine the joint bandwidth and computing resource reservation. The proposed UDT-assisted resource reservation scheme can be applied to analyze the user's behavior pattern and integrate its impact on resource management in interactive media scenarios. For future work, we will investigate the joint optimization of the segment-level caching order and resource allocation based on distilled information from UDTs to further improve user satisfaction.


\bibliographystyle{IEEEtran}
\bibliography{ref}

\begin{thebibliography}{10}
\providecommand{\url}[1]{#1}
\csname url@samestyle\endcsname
\providecommand{\newblock}{\relax}
\providecommand{\bibinfo}[2]{#2}
\providecommand{\BIBentrySTDinterwordspacing}{\spaceskip=0pt\relax}
\providecommand{\BIBentryALTinterwordstretchfactor}{4}
\providecommand{\BIBentryALTinterwordspacing}{\spaceskip=\fontdimen2\font plus
\BIBentryALTinterwordstretchfactor\fontdimen3\font minus
  \fontdimen4\font\relax}
\providecommand{\BIBforeignlanguage}[2]{{%
\expandafter\ifx\csname l@#1\endcsname\relax
\typeout{** WARNING: IEEEtran.bst: No hyphenation pattern has been}%
\typeout{** loaded for the language `#1'. Using the pattern for}%
\typeout{** the default language instead.}%
\else
\language=\csname l@#1\endcsname
\fi
#2}}
\providecommand{\BIBdecl}{\relax}
\BIBdecl

\bibitem{prework}
X.~Huang, W.~Wu, and X.~Shen, ``Digital twin-assisted resource demand
  prediction for multicast short video streaming,'' in \emph{Proc. IEEE Int.
  Conf. Distrib. Comput. Syst. (ICDCS) PhD Stu. Symp.}, Hong Kong, China, 2023.

\bibitem{yuan2019spatial}
H.~Yuan, S.~Zhao, J.~Hou, X.~Wei, and S.~Kwong, ``Spatial and temporal
  consistency-aware dynamic adaptive streaming for 360-degree videos,''
  \emph{IEEE J. Sel. Top. Signal Process.}, vol.~14, no.~1, pp. 177--193, 2019.

\bibitem{wang2022task}
K.~Wang, J.~Jin, Y.~Yang, T.~Zhang, A.~Nallanathan, C.~Tellambura, and
  B.~Jabbari, ``Task offloading with multi-tier computing resources in next
  generation wireless networks,'' \emph{IEEE J. Sel. Areas Commun.}, vol.~41,
  no.~2, pp. 306--319, 2022.

\bibitem{report}
\BIBentryALTinterwordspacing
{Buiness of Apps}, ``Tiktok revenue and usage statistics 2023,'' 2023.
  [Online]. Available:
  \url{https://www.businessofapps.com/data/tik-tok-statistics/}
\BIBentrySTDinterwordspacing

\bibitem{panor}
W.~Zhang, F.~Qian, B.~Han, and P.~Hui, ``Deepvista: {16K} panoramic cinema on
  your mobile device,'' in \emph{Proc. Int. World Wide Web (WWW)}, Ljubljana,
  Slovenia, 2021, pp. 2232--2244.

\bibitem{li2023utility}
J.~Li, Y.~Xu, Y.~Cao, J.~Zhu, and D.~Wang, ``Utility-driven joint caching and
  bitrate allocation for real-time immersive videos,'' \emph{IEEE J. Sel. Top.
  Signal Process.}, 2023.

\bibitem{Multicast}
K.~Zahoor, K.~Bilal, A.~Erbad, and A.~Mohamed, ``Service-less video multicast
  in {5G}: Enablers and challenges,'' \emph{IEEE Netw.}, vol.~34, no.~3, pp.
  270--276, 2020.

\bibitem{9676649}
K.~Wang, W.~Chen, J.~Li, Y.~Yang, and L.~Hanzo, ``Joint task offloading and
  caching for massive {MIMO}-aided multi-tier computing networks,'' \emph{IEEE
  Trans. Commun.}, vol.~70, no.~3, pp. 1820--1833, 2022.

\bibitem{yans}
N.~Reyhanian and Z.-Q. Luo, ``Data-driven adaptive network slicing for
  multi-tenant networks,'' \emph{IEEE J. Sel. Top. Signal Process.}, vol.~16,
  no.~1, pp. 113--128, 2021.

\bibitem{xinyu_transcoding}
X.~Huang, M.~Li, W.~Wu, C.~Zhou, and X.~Shen, ``Digital twin-assisted
  collaborative transcoding for better user satisfaction in live streaming,''
  in \emph{Proc. IEEE Int. Conf. Commun. (ICC)}, Rome, Italy, 2023.

\bibitem{reservation}
X.~Tan, L.~Xu, Q.~Zheng, S.~Li, and B.~Liu, ``{QoE}-driven {DASH} multicast
  scheme for {5G} mobile edge network,'' \emph{J. Commun. Inf. Netw.}, vol.~6,
  no.~2, pp. 153--165, 2021.

\bibitem{hebo}
B.~He, J.~Wang, Q.~Qi, H.~Sun, H.~Zhou, L.~Zhang, K.~Liu, and J.~Liao,
  ``Resilient {QUIC} protocol for emerging wireless networks,'' \emph{IEEE
  Wirel. Commun.}, vol.~29, no.~3, pp. 64--70, 2022.

\bibitem{zhu2022swipe}
S.~Zhu, T.~Karagioules, E.~Halepovic, A.~Mohammed, and A.~D. Striegel, ``Swipe
  along: a measurement study of short video services,'' in \emph{Proc. ACM
  Multimedia Syst. Conf.}, Athlone, Ireland, 2022, pp. 123--135.

\bibitem{digital}
M.~Grieves, ``Digital twin: manufacturing excellence through virtual factory
  replication,'' \emph{White paper}, vol.~1, no. 2014, pp. 1--7, 2014.

\bibitem{xu2019optimal}
W.~Xu, Y.~Cui, and Z.~Liu, ``Optimal multi-view video transmission in multiuser
  wireless networks by exploiting natural and view synthesis-enabled multicast
  opportunities,'' \emph{IEEE Trans. Commun.}, vol.~68, no.~3, pp. 1494--1507,
  2019.

\bibitem{li2018performance}
M.~Li and Y.-H. Wu, ``Performance analysis of adaptive multicast streaming
  services in wireless cellular networks,'' \emph{IEEE Trans. Mobile Comput.},
  vol.~18, no.~11, pp. 2616--2630, 2018.

\bibitem{zahoor2020service}
K.~Zahoor, K.~Bilal, A.~Erbad, and A.~Mohamed, ``Service-less video multicast
  in {5G}: Enablers and challenges,'' \emph{IEEE Netw.}, vol.~34, no.~3, pp.
  270--276, 2020.

\bibitem{Zhang-mult}
X.~Zhang, M.~Yang, Y.~Zhao, J.~Zhang, and J.~Ge, ``An {SDN}-based video
  multicast orchestration scheme for {5G} ultra-dense networks,'' \emph{IEEE
  Commun. Mag.}, vol.~55, no.~12, pp. 77--83, 2017.

\bibitem{10077734}
N.~A. Mitsiou, V.~K. Papanikolaou, P.~D. Diamantoulakis, T.~Q. Duong, and G.~K.
  Karagiannidis, ``Digital twin-aided orchestration of mobile edge computing
  with grant-free access,'' \emph{IEEE Open J. Commun. Soc.}, vol.~4, pp.
  841--853, 2023.

\bibitem{NOMA-group}
J.~Montalban, P.~Scopelliti, M.~Fadda, E.~Iradier, C.~Desogus, P.~Angueira,
  M.~Murroni, and G.~Araniti, ``Multimedia multicast services in {5G} networks:
  Subgrouping and non-orthogonal multiple access techniques,'' \emph{IEEE
  Commun. Mag.}, vol.~56, no.~3, pp. 91--95, 2018.

\bibitem{soni2017nfv}
H.~Soni, W.~Dabbous, T.~Turletti, and H.~Asaeda, ``{NFV}-based scalable
  guaranteed-bandwidth multicast service for software defined isp networks,''
  \emph{IEEE Trans. Netw. Serv. Manage.}, vol.~14, no.~4, pp. 1157--1170, 2017.

\bibitem{qiao2018improving}
S.~Mandal, G.~Maji, S.~Khatua, and R.~K. Das, ``Cost minimizing reservation and
  scheduling algorithms for public clouds,'' \emph{IEEE Trans. Cloud Comput.},
  vol.~11, no.~2, pp. 1365--1380, 2021.

\bibitem{araniti2018hybrid}
G.~Araniti, P.~Scopelliti, G.-M. Muntean, and A.~Iera, ``A hybrid
  unicast-multicast network selection for video deliveries in dense
  heterogeneous network environments,'' \emph{IEEE Trans. Broadcast.}, vol.~65,
  no.~1, pp. 83--93, 2018.

\bibitem{ye2018network}
Q.~Ye, J.~Li, K.~Qu, W.~Zhuang, X.~Shen, and X.~Li, ``A network slicing
  framework for end-to-end {QoS} provisioning in {5G} networks,'' \emph{IEEE
  Veh. Technol. Mag.}, vol.~13, no.~2, pp. 65--74, 2018.

\bibitem{xu2021tripres}
X.~Xu, Z.~Fang, L.~Qi, X.~Zhang, Q.~He, and X.~Zhou, ``Tripres: Traffic flow
  prediction driven resource reservation for multimedia {IoV} with edge
  computing,'' \emph{ACM Trans. Multimedia Comput. Commun. Appl.}, vol.~17,
  no.~2, pp. 1--21, 2021.

\bibitem{glaessgen2012digital}
E.~Glaessgen and D.~Stargel, ``The digital twin paradigm for future {NASA} and
  {U.S.} air force vehicles,'' in \emph{Proc. Struct. Dyn. Mater. Conf. Special
  Session: Digital Twin}, Honolulu, HI, USA, 2012, p. 1818.

\bibitem{holi}
X.~Shen, J.~Gao, W.~Wu, M.~Li, C.~Zhou, and W.~Zhuang, ``Holistic network
  virtualization and pervasive network intelligence for {6G},'' \emph{{IEEE}
  Commun. Surveys Tuts.}, vol.~24, no.~1, pp. 1--30, 2022.

\bibitem{10183802}
T.~Q. Duong, D.~Van~Huynh, S.~R. Khosravirad, V.~Sharma, O.~A. Dobre, and
  H.~Shin, ``From digital twin to metaverse: The role of {6G} ultra-reliable
  and low-latency communications with multi-tier computing,'' \emph{IEEE Wirel.
  Commun.}, vol.~30, no.~3, pp. 140--146, 2023.

\bibitem{9939166}
A.~Masaracchia, V.~Sharma, B.~Canberk, O.~A. Dobre, and T.~Q. Duong, ``Digital
  twin for {6G}: Taxonomy, research challenges, and the road ahead,''
  \emph{IEEE Open J. Commun. Soc.}, vol.~3, pp. 2137--2150, 2022.

\bibitem{sun2020reducing}
W.~Sun, H.~Zhang, R.~Wang, and Y.~Zhang, ``Reducing offloading latency for
  digital twin edge networks in {6G},'' \emph{IEEE Trans. Veh. Technol.},
  vol.~69, no.~10, pp. 12\,240--12\,251, 2020.

\bibitem{van2022urllc}
D.~Van~Huynh, V.-D. Nguyen, S.~R. Khosravirad, V.~Sharma, O.~A. Dobre, H.~Shin,
  and T.~Q. Duong, ``{URLLC} edge networks with joint optimal user association,
  task offloading and resource allocation: A digital twin approach,''
  \emph{IEEE Trans. Commun.}, vol.~70, no.~11, pp. 7669--7682, 2022.

\bibitem{liu2023energy}
W.~Liu, B.~Li, W.~Xie, Y.~Dai, and Z.~Fei, ``Energy efficient computation
  offloading in aerial edge networks with multi-agent cooperation,'' \emph{IEEE
  Trans. Wirel. Commun.}, 2023.

\bibitem{peng2022distributed}
K.~Peng, H.~Huang, M.~Bilal, and X.~Xu, ``Distributed incentives for
  intelligent offloading and resource allocation in digital twin driven smart
  industry,'' \emph{IEEE Trans. Industr. Inform.}, vol.~19, no.~3, pp.
  3133--3143, 2022.

\bibitem{zheng2023data}
J.~Zheng, T.~H. Luan, Y.~Zhang, R.~Li, Y.~Hui, L.~Gao, and M.~Dong, ``Data
  synchronization in vehicular digital twin network: A game theoretic
  approach,'' \emph{IEEE Trans. Wirel. Commun.}, 2023.

\bibitem{lu2021adaptive}
Y.~Lu, S.~Maharjan, and Y.~Zhang, ``Adaptive edge association for wireless
  digital twin networks in {6G},'' \emph{IEEE Internet Things J.}, vol.~8,
  no.~22, pp. 16\,219--16\,230, 2021.

\bibitem{zhou2022}
C.~Zhou, J.~Gao, M.~Li, X.~Shen, and W.~Zhuang, ``Digital twin-empowered
  network planning for multi-tier computing,'' \emph{J. Commun. Inf. Netw.},
  vol.~7, no.~3, pp. 221--238, 2022.

\bibitem{hu-traffic}
C.~Hu, W.~Fan, E.~Zeng, Z.~Hang, F.~Wang, L.~Qi, and M.~Z.~A. Bhuiyan,
  ``Digital twin-assisted real-time traffic data prediction method for
  5g-enabled internet of vehicles,'' \emph{IEEE Trans. Industr. Infor.},
  vol.~18, no.~4, pp. 2811--2819, 2022.

\bibitem{vaezi2022digital}
M.~Vaezi, K.~Noroozi, T.~D. Todd, D.~Zhao, G.~Karakostas, H.~Wu, and X.~Shen,
  ``Digital twins from a networking perspective,'' \emph{IEEE Internet Things
  J.}, vol.~9, no.~23, pp. 23\,525--23\,544, 2022.

\bibitem{li2023dashlet}
Z.~Li, Y.~Xie, R.~Netravali, and K.~Jamieson, ``Dashlet: Taming swipe
  uncertainty for robust short video streaming,'' in \emph{Proc. USENIX Symp.
  Netw. Syst. Des. Implement. (NSDI)}, Boston, MA, USA, 2023, pp. 1583--1599.

\bibitem{G_Huang}
G.~Huang, W.~Gong, B.~Zhang, C.~Li, and C.~Li, ``An online buffer-aware
  resource allocation algorithm for multiuser mobile video streaming,''
  \emph{IEEE Trans. Veh. Technol.}, vol.~69, no.~3, pp. 3357--3369, 2020.

\bibitem{Gong}
X.~Gong, Q.~Feng, Y.~Zhang, J.~Qin, W.~Ding, B.~Li, P.~Jiang, and K.~Gai,
  ``Real-time short video recommendation on mobile devices,'' in \emph{Proc.
  ACM Int. Conf. Infor. Knowl. Management}, Atlanta, GA, USA, 2022, pp.
  3103--3112.

\bibitem{kmeans}
D.~Arthur and S.~Vassilvitskii, ``k-means++: The advantages of careful
  seeding,'' Stanford, Tech. Rep., 2006.

\bibitem{wenwen}
W.~Jiang, B.~Ai, J.~Cheng, Y.~Lin, and G.~Zhang, ``Sum of age-of-information
  minimization in aerial irss assisted wireless networks,'' \emph{IEEE Commun.
  Lett.}, vol.~27, no.~5, pp. 1377--1381, 2023.

\bibitem{DDQN}
H.~Van~Hasselt, A.~Guez, and D.~Silver, ``Deep reinforcement learning with
  double q-learning,'' in \emph{Proc. {AAAI} Conf. Artif. Intell.}, vol.~30,
  no.~1, Phoenix, Arizona, USA, 2016.

\bibitem{Autoencoder}
A.~Ng, ``Sparse autoencoder,'' \emph{CS294A Lecture notes}, vol.~72, no. 2011,
  pp. 1--19, 2011.

\bibitem{Globe}
X.~Huang, C.~Zhou, W.~Wu, M.~Li, H.~Wu, and X.~Shen, ``Personalized {QoE}
  enhancement for adaptive video streaming: A digital twin-assisted scheme,''
  in \emph{Proc. IEEE Glob. Commun. Conf. (GLOBECOM)}, 2022, pp. 4001--4006.

\bibitem{zhang2020satisfied}
X.~Zhang, C.~Yang, H.~Wang, W.~Xu, and C.-C.~J. Kuo, ``Satisfied-user-ratio
  modeling for compressed video,'' \emph{IEEE Trans. Image Process.}, vol.~29,
  pp. 3777--3789, 2020.

\bibitem{Wen}
W.~Wu, N.~Chen, C.~Zhou, M.~Li, X.~Shen, W.~Zhuang, and X.~Li, ``Dynamic {RAN}
  slicing for service-oriented vehicular networks via constrained learning,''
  \emph{IEEE J. Sel. Areas Commun.}, vol.~39, no.~7, pp. 2076--2089, 2021.

\bibitem{convex}
D.~Bertsekas, \emph{Network optimization: continuous and discrete
  models}.\hskip 1em plus 0.5em minus 0.4em\relax Athena Scientific, 1998,
  vol.~8.

\bibitem{boyd}
S.~Boyd, S.~P. Boyd, and L.~Vandenberghe, \emph{Convex optimization}.\hskip 1em
  plus 0.5em minus 0.4em\relax Cambridge university press, 2004.

\bibitem{DBSCAN}
E.~Schubert, J.~Sander, M.~Ester, H.~P. Kriegel, and X.~Xu, ``{DBSCAN}
  revisited, revisited: why and how you should (still) use {DBSCAN},''
  \emph{ACM Trans. Database Syst.}, vol.~42, no.~3, pp. 1--21, 2017.

\bibitem{BBM}
V.~I. Norkin, G.~C. Pflug, and A.~Ruszczy{\'n}ski, ``A branch and bound method
  for stochastic global optimization,'' \emph{Math. Program.}, vol.~83, pp.
  425--450, 1998.

\bibitem{BDQ}
A.~Tavakoli, F.~Pardo, and P.~Kormushev, ``Action branching architectures for
  deep reinforcement learning,'' in \emph{Proc. {AAAI} Conf. Artif. Intell.},
  vol.~32, no.~1, 2018.

\end{thebibliography}

\end{document}